\documentclass[usegraphicx,usenatbib,useapjfonts,apj]{emulateapj}

\def\lesssim{{_ <\atop{^\sim}}}

\def\fb{\mbox{$f_{\rm gal}$}}
\def\fg{\mbox{$f_{\rm g}$}}
\def\fs{\mbox{$f_{\rm s}$}}
\def\kms{\mbox{kms$^{-1}$}}

\def\hb{\mbox{$R_{\rm bar}$}}
\def\hB{\mbox{$R_{\rm B}$}}
\def\hK{\mbox{$R_{\rm K}$}}
\def\hs{\mbox{$R_{\rm s}$}}

\def\mv{\mbox{$M_{\rm vir}$}}
\def\rv{\mbox{$R_{\rm vir}$}}
\def\rh{\mbox{$R_{\rm h}$}}
\def\md{\mbox{$M_{\rm bar}$}}
\def\mg{\mbox{$M_{\rm g}$}}
\def\ms{\mbox{$M_{\rm s}$}}
\def\mh{\mbox{$M_{\rm h}$}}
\def\msun{\mbox{M$_\odot$}}

\def\sigd{\mbox{$\Sigma_{\rm bar,0}$}}
\def\sigb{\mbox{$\Sigma_{\rm bar,0}$}}

\def\sigB{\mbox{$\Sigma_{\rm B,0}$}}
\def\sigK{\mbox{$\Sigma_{\rm K,0}$}}
\def\sigs{\mbox{$\Sigma_{\rm s,0}$}}

\def\muK{\mbox{$\mu_{\rm K,0}$}}

\def\vm{\mbox{$V_{\rm m}$}}
\def\vmh{\mbox{$V_{\rm m,h}$}}

\def\W{\mbox{$W_{20}$}}
\def\LCDM{\mbox{$\Lambda$CDM}}

\def\msunh{\mbox{M$_{\odot}h^{-1}$}}

\def\sigintr{\mbox{$\sigma_{\rm intr}$}}
\def\sigintrV{\mbox{$\sigma_{\rm intr}({\rm Log\vm})$}}

\def\sigintrR{\mbox{$\sigma_{\rm intr}({\rm LogR})$}}

\def\sigobs{\mbox{$\sigma_{\rm obs}$}}
\def\sigfit{\mbox{$\sigma_{\rm fit}$}}
\def\sigVMh{\mbox{$\sigma_{\rm MAH}$(Log\vmh)}}

\slugcomment{To appear in Astronomical Journal}

\shorttitle{Scaling relations of disk galaxies}
\shortauthors{Avila-Reese et al.}

\begin{document}

\title{On the baryonic, stellar, and luminous scaling relations of disk galaxies}

\author{V. Avila-Reese\altaffilmark{1}, J. Zavala\altaffilmark{2,3}, C. Firmani\altaffilmark{4,1},
and H. M. Hern\'andez-Toledo\altaffilmark{1}}
 





\altaffiltext{1}{Instituto de Astronom\'{\i}a, Universidad Nacional Aut\'onoma de M\'exico, A.P. 70-264, 04510 M\'exico D. F., M\'exico}
\altaffiltext{2}{Instituto de Ciencias Nucleares, Universidad Nacional Aut\'onoma de M\'exico, A.P. 70-543, 04510 M\'exico D.F., M\'exico.}
\altaffiltext{3}{{\it Present affiliation:} Shanghai Astronomical Observatory, Nandan Road 80, Shanghai 200030, China; jzavala@shao.ac.cn}
\altaffiltext{4}{INAF--Osservatorio Astronomico di Brera, via E.Bianchi 46, I--23807 Merate, Italy}



\begin{abstract}

We explore how the slopes and scatters of the scaling relations of disk galaxies 
(\vm-$L$[-$M$], $R$-$L$[-$M$], and \vm-$R$) do change when moving from $B$ to $K$ bands 
and to stellar and baryonic quantities.  For our compiled sample of 76 normal, 
non--interacting high and low surface brightness (SB) disk galaxies, we find important changes, 
which evidence evolution effects, mainly related to the gas infall and star 
formation (SF) processes. We also explore correlations among the $(B-K)$ color, stellar mass 
fraction \fs, mass $M$ (luminosity $L$), and surface density (SB), as well as correlations 
among the residuals of the scaling relations, and among these residuals and those of the 
other relations studied here. Some of our findings are: (i) the scale length \hb~ is a third 
parameter in the baryonic Tully--Fisher relation (TFR) and the residuals of this relation 
follow a trend (slope $\approx -0.15$) with the residuals of the \hb-\md~ relation; for 
the stellar and $K$ band cases, the scale length is not anymore a third parameter and the 
mentioned trend disappears; (ii) among the TFRs, the $B-$band TFR is the most scattered; 
in this case, the color is a third parameter, in agreement with previous works; (iii) the 
low SB galaxies break some observed trends in diagrams that include surface density, 
color, and \fs, suggesting then a threshold in the gas surface density
$\Sigma_g$, below which the SF becomes independent of the gas infall rate and $\Sigma_g$. 
Our results are interpreted and discussed in the light of $\Lambda$ Cold Dark Matter--based 
models of disk--galaxy 
formation and evolution. These models are able to explain not only the baryonic 
scaling correlations, but also most of the processes responsible for the observed changes in the
slopes, scatters, and correlations among the residuals when changing to stellar and luminous 
quantities.  The galaxy baryon fraction, \fb, is required to be smaller than 0.05
on average. We detect some potential difficulties for the models: the observed 
color-$M$ and surface density-$M$ correlations are steeper, and the intrinsic scatter 
in the baryonic TFR is smaller than those predicted.

\end{abstract}

\section{Introduction}

Disk galaxies are the main population of galaxies in the local Universe.
The study of their global scaling relations is of paramount relevance for
understanding the formation and evolution of galaxies in general.  The main 
disk--galaxy observed global scaling relations are those between maximum circular 
velocity, \vm, and luminosity, $L$; between disk scale length, $R$, and 
$L$; and between \vm\ and $R$. Each one of these relations can be 
established for different optical/near--IR (NIR) pass--bands. Under some assumptions,
analogous relations can be calculated for the corresponding stellar and 
baryonic quantities. The changes in the zero points, slopes and scatters of 
these relations, when changing from one color band to another, and to stellar and
baryonic quantities, are expected to be intimately related to the stellar--population 
properties and radial distributions of disk galaxies. In turn, these properties and 
distributions depend mainly on the disk assembly and star--formation (hereafter SF) 
histories. Therefore, the study of the scale relations 
for different passbands and for the stellar and baryonic cases offers valuable 
information on the structural, dynamical and SF properties of disk 
galaxies, as well as important constraints on models of their formation and evolution.
The present paper is focused on this study.

\subsection{Background on the disk scaling relations}

The most robust and studied of the disk--galaxy scaling relations is the one
between $L$ and \vm, the so--called Tully--Fisher relation (TFR).
Originally discovered in the $B$ passband (Tully \& Fisher 1977), it was 
afterward shown that it also applies to infrared bands. More recently, after 
applying a velocity (luminosity)--dependent extinction correction, Tully \& Pierce 
(2000) have confirmed that the slope of the $L$-$\vm$ correlation gets steeper 
systematically from $B$ ($\approx 2.9$) to $K$ ($\approx 3.5$) bands.
This result suggests some dependence 
of galaxy colors on \vm\ for a given galaxy luminosity or mass.
The amplitude of the scatter of the TFRs changes also with wavelength.
These dependencies 
of the slope and scatter of the TFR on wavelength are related mainly to stellar 
population effects, which on their own, are related to the assembly and SF
histories of galaxies.  Regarding the intrinsic TFR scatters, they are found 
to be small, posing a challenge for scenarios aimed to explain the origin 
of disk galaxies (Eisenstein \& Loeb 1996; Avila-Reese et al. 1998; 
Mo et al. 1998; Firmani \& Avila-Reese 2000).

The TFR has widely been studied for estimating extragalactic distances.
In this case the independent variable (predictor) is the circular 
velocity, and the galaxy samples are carefully pruned to minimize the scatter. 
The latter implies a selection of samples for 
a narrow range of morphological types and surface brightnesses (hereafter SBs). 
However, for studies that aim to understand the origin of the TFR and use it to 
constrain models of galaxy formation rather than circular velocity, the 
luminosity should be the predictor, and the sample should be as wide as possible 
in morphological types, SBs, and colors (Courteau et al. 2007). Hereafter, we will 
refer to the TFR as the \vm-$L$(-$M$) correlation.

By using NIR magnitudes --which are good tracers of stellar masses-- 
and gas contents when available, some authors have calculated the stellar and baryonic TFRs 
(e.g., McGaugh et al. 2000; Bell \& de Jong 2001; Mayer \& Moore 2004; Gurovich et al. 
2004; Pizagno et al. 2005; de Rijcke et al. 2007). As discussed in Firmani \& Avila-Reese 
(2000; hereafter FA-R), on one hand the baryonic TFR appears to be closely 
related to the cosmological halo maximum circular velocity--mass relation, and on 
the other hand, when compared to the stellar TFR, it 
reflects some aspects of the SF efficiency for different types 
of galaxies. The behavior of the scatter, when one changes from $B-$ to 
$K-$band TFRs and to the stellar and baryonic TFRs, certainly contains 
relevant information on the processes of disk galaxy evolution such as 
gas infall, SF, and stellar population histories.

The radius for a given luminosity (or the disk SB) has been considered 
as a potential third parameter that correlates with the scatter of the TFR 
(e.g., Kodaira 1989; Willick 1999). The scale length $R$ is indeed 
another structural parameter of disk galaxies that span a wide range of 
values. Some authors have reported that the scatter of the TFR is reduced 
up to 50\% when $R$ is introduced as a third parameter 
(e.g., Koda, Sofue \& Wada 2000; Han et al. 2001).  However, 
if the scatter of a given relation decreases upon introducing a third variable, it
does not necessarily mean that such a variable is a statistically significant 
parameter. Indeed, some authors have shown that $R$ (or central SB) is
not a physical third parameter in the TFR (e.g., Han 1991; Tully \& Verheijen 1997; 
Willick et al. 1997; Courteau \& Rix 1999; FA-R), but the radius could be a third 
parameter in the baryonic TFR (FA-R; Avila-Reese, Firmani \& Zavala 2002; 
Zavala 2003; Dutton et al. 2007). 

It is well known that the radius correlate with $L$ 
(Freeman 1970) and \vm\ (Tully \& Fisher 1977), although these correlations 
are much more scattered than the TFR. Disk galaxies in the 
Log$L$--Log$\vm$--Log$R$ parameter--space indeed lie in a thin oblique plane 
whose edge--on projection is close to the Log$L-$Log\vm\ (TFR) projection 
(Koda, Sofue \& Wada 2000; Han et al. 2001; Shen, Mo \& Shu 2002; Gnedin
et al. 2007). This is the so called fundamental plane of disk galaxies. Here we 
construct from observations and analyze separately each one of 
the three projections of these planes.

\subsection{Construction and interpretation of the scaling relations:
how do they change from luminous to stellar and baryonic quantities?}

In most of the previous works, the disk--galaxy scaling correlations were constructed 
for samples biased in morphological types and SBs (pruned for reducing the scatter), 
and without selecting the environment. However, {\it for theoretical and interpretative 
studies, the  observational galaxy sample should be as wide as possible in the range of 
morphologies, SBs, and colors} in order to include the whole population of disk galaxies, 
later to be compared with model predictions.
Besides, since galaxy models are commonly used for isolated galaxies, the observational 
sample should also be selected to contain galaxies {\it not affected by the environment}.
On the other hand, the limited information of the samples employed in previous 
works, did not allow one to explore how the scaling correlations and their scatters 
change from luminous to stellar and baryonic quantities.

In this paper we shall use a sample of $76$ disk galaxies compiled from
the literature and homogenized in Zavala et al. (2003a; hereafter ZAHF)
and Zavala (2003).  This sample is not in any sense complete, 
an aspect that is crucial for volume--averaged population studies (e.g., the
luminosity function), but is less relevant for studies on the correlations
of physical properties. The sample includes high and low SB 
(hereafter HSB and LSB, respectively) disk {\it normal}\footnote{Here we understand 
by normal galaxies: non--dwarf, non--interacting, nearly isolated local 
galaxies.} galaxies of all morphological types, and it provides
photometric information, at least in $B$ and $K$ bands, as well as information
on the rotation curve and the HI integrated flux. 
None of the previous related works have presented a sample with these
characteristics, which are crucial for the analysis proposed here.
By using this sample, we will explore the differences in the slope and scatter   
of each scaling relation defined in the $B$ and $K$ bands, and for stellar and 
baryonic quantities. The results of this exploration will be 
interpreted in the light of disk--galaxy formation and evolution models.

The current theory of galaxy formation and evolution is based on the 
hierarchical $\Lambda$ Cold Dark Matter (\LCDM) cosmological model 
(for recent reviews, see Baugh 2006; Avila-Reese 2007). Several  authors have attempted
to predict the scaling relations of disk galaxies and their scatters
within this model (e.g., Avila-Reese et al. 1998,2002; Mo et al. 1998; 
Steinmetz \& Navarro 1999; Navarro \& Steinmetz 2000; FA-R; Koda et al. 2000; 
Shen et al. 2002; Zavala 2003; Courteau et al. 2003,2007; Dutton et al. 
2005,2007; Gnedin et al. 2007; Courteau et al. 2007). 
The results are encouraging. We shall use some of these results, mainly those
of FA-R, to interpret and discuss the new inferences from observations presented here. 

Note that it was a common practice in the literature to compare 
model results with the observed {\it luminous} properties and
correlations, while most of the galaxy models did not actually include or 
self--consistently treat the process of gas transformation into stars, as well as did 
not follow the evolutionary path of halos and galaxies. As FA-R showed (see also 
Avila-Reese et al. 2002; Dutton et al. 2005), on one hand, there are systematical 
differences in the predicted scaling relations and their scatters among the luminous, 
stellar, and baryonic quantities; differences that we will explore here. On
the other hand, properties such as the stellar and baryonic $M$, $R$, \vm, gas
fraction, and integral colors depend not only on the initial parameters, but 
also on the evolutionary path of the halo--disk system (Stringer \& Benson 2008). 
Contrary to the ``static population'' models, such as those of Mo et al. (1998) 
and extensions of them, the
FA-R models are evolutionary. With some simplifications, they follow self--consistently
(see Appendix B): the halo and disk assembly, taking into account the halo adiabatic 
contraction; the self--regulated disk SF and feedback; the luminosity evolution; and the 
secular bulge formation.

During the completion of the present work, several of the papers mentioned above, 
with results and a philosophy similar to ours, appeared in the literature. 
Although some overlapping is present,
our contribution is original in the sense that we focus the research {\it on the 
changes in the scaling relations and their scatters when moving from optical to 
NIR bands and to stellar and baryonic quantities}.  Our sample is similar in number to 
others used in recent works (e.g., Gnedin et al. 2007) and smaller with respect
to Dutton et al. (2007) and Courteau et al. (2007), who use $\sim 1300$ objects with 
$I-$ band photometry and $H\alpha$ kinematics, but only $\sim 360$ out of them, mainly 
HSB galaxies, 
have available $K-$ band photometry. These large samples are undoubtedly useful 
and we will compare, when possible, with their results. However, we can not use them 
because they lack several observables (e.g., gas content, optical photometry) 
and requirements (e.g., inclination limits, LSB galaxies in both optical and $K$ bands) 
that we have imposed for our study. We hope that our study will serve as a benchmark for 
similar analyzes based on large multiwavelength galaxy surveys, which are currently 
being completed.  

This paper is organized as follows. In \S 2 we describe the disk
galaxy sample to be used. The three main scaling 
correlations and their scatters are constructed and discussed in \S 3 for bands 
$B$ and $K$ and for the inferred stellar and baryonic quantities. Some other
correlations among global galaxy quantities are also reported. In \S 4, in 
the light of previous \LCDM--based disk galaxy formation and evolution models,  
we interpret the changes in the slopes and scatters of the scaling relations
when changing from $B$ to $K$ bands and to stellar and baryonic quantities.
A summary of the results and the conclusions of this work are presented in \S 6.

\section{The sample} 

In the literature there are only a few small galaxy samples with the requirements 
and data we need (broad in morphological types and SBs, with information on 
detailed surface--photometry in two separated pass--bands, e.g. $B$ and $K$, on 
global dynamics,  and on total HI gas flux). Here we use a compilation and 
completion of these samples presented in ZAHF and Zavala (2003; see also Graham 2002). 
In the following, we present 
the sub--samples, describe the homogenization and data correction 
processes that we have implemented, and the way in which we calculated 
the stellar and baryonic quantities (see ZAHF for additional details).

The sub--samples are: (i) de Jong \& van der Kruit (1994; see 
also de Jong 1996a), which comprises undisturbed field, mostly HSB, disk 
galaxies, (ii) Verheijen (1997; see also Verheijen \& Sancisi 2001), composed
of disk HSB and LSB galaxies from the least massive and most
spiral--rich nearby cluster, Ursa Minor,  and (iii) Bell et al. (2000),
a sample of LSB galaxies observed partially by them; the rest were compiled 
from the literature, mainly from de Blok et al. (1995,1996).  In our 
sample we also included the Milky Way and Andromeda galaxies.

The raw photometric magnitudes, disk scale lengths, $R$, and central SBs 
were taken directly from the source papers. In ZAHF and 
Zavala (2003) we have checked that there were no significant systematic 
differences in these parameters with inclination; we also found no systematics
due to the different completeness criteria and profile fitting procedures used 
by the original authors of each sub--sample. 
In fact, the dominant source of error in the photometric parameters is 
caused by the uncertainty in the sky level (Bell et al 2000; de Jong 1996a).
Regarding the dynamical information, and to make the data as uniform as possible, 
we used the velocity line--width \W\ (defined at the 20\% level) for almost all 
the galaxies in the sample. Only a few (LSB) galaxies in the sample lack a \W\ 
measurement but have measured rotation curves; in these cases we have
used the \vm\ inferred directly from the rotation curve. If properly
corrected, \W\ $\approx 2\vm$ (Verheijen 1997; Verheijen 2001). 
For the Verheijen sub--sample, we adopt the \W\
values given in Verheijen \& Sancisi (2001), while for the de Jong sub--sample 
and most of the LSB galaxies from the Bell et al. sub--sample, \W\ was extracted 
from the HyperLEDA information system\footnote{http://leda.univ-lyon1.fr}.  
The integral fluxes in the 21--cm 
line (used to calculate the neutral gas mass component in galaxies) were taken from 
HyperLEDA, Verheijen \& Sancisi (2001), and de Blok et al. (1996).

A condition imposed on our sample is that galaxies should be in a 
restricted range of inclinations ($35^{o} \leq i \leq 80^{o}$).
We excluded from the sample any galaxies with clear signs of interaction
(mainly from the Verjeihen sub--sample) and with rotation curves that are
still increasing at the last measured outer radius. Only some LSB galaxies
showed this feature. 
The final sample consists of 76 galaxies: 42 out of 86, 29 out of 52,
and 5 (8) out of 23 from the de Jong (1996a), Verheijen (1997), and Bell et al. 
(2000) sub--samples,
respectively, plus the Milky Way and Andromeda galaxies. From the Bell et al. 
(2000) sample, 8 LSB galaxies are useful for us, but three of them 
are in common with the de Jong sample.  

The local distances to the galaxies were calculated using the kinematic
distance modulus given in HyperLEDA (see more details in ZAHF). The value of 
the Hubble constant used in this database is the same as that we assume here,
$H_{0}=70 \kms$Mpc$^{-1}$. Since the LSB galaxies from the Bell et al. (2000)
sub--sample are not included in HyperLEDA, their distances were taken directly 
from the source paper, properly corrected for the H$_{0}$ value used here. 

The compiled sample is not complete in any sense, but it 
comprises representative ranges of the basic features of {\it normal} 
disk galaxies (see Fig. 1 in ZAHF). These features are mainly
the morphological type, magnitude, disk central SB, integral color, and 
gas fraction. 

\subsection{Data corrections}

The total magnitudes were corrected for Galactic extinction (following Schlegel 
et al. 1998), $K$ term, and internal extinction (see ZAHF for details). 
For the latter we used the empirical velocity-- (luminosity--) 
dependent extinction coefficients determined by Tully et al. (1998).  
The central SBs were corrected for Galactic extinction, $K$ term,
cosmological SB dimming, and inclination --geometrical and extinction
effects-- (see ZAHF). For the inclination correction, we followed Verheijen (1997),
considering LSB galaxies optically thin in all bands. The HSB galaxies were also 
considered optically thin in the $K$ band. We have defined LSB galaxies as those 
with central SBs in the $K$ band, after SB correction, larger than 18.5 mag/arcsec$^2$ 
(Verheijen 1997). The 21 cm line--widths at the 20\% level, \W, for the Verheijen 
susbsample were taken directly from  Verheijen \& Sancisi (2001). Galaxies with
\W\ taken from LEDA were de--corrected (for instrumental corrections) to get 
the raw data (Paturel et al. 1997), and then corrected again for broadening 
due to turbulent motions and for inclination, this time following the
procedure by Verheijen \& Sancisi (2001).

In ZAHF, we did not take into account the observational errors. Here we
attempt to estimate these errors for both the photometric and dynamic parameters,
and use them for our results when possible. In Appendix A, we present our approach to
estimate the errors.

\subsection{Composite quantities}

The stellar mass and disk central surface density, \ms\ and \sigs, were
derived from the $K-$band luminosity (which includes the bulge) and the 
extrapolated disk central SB, \sigK, respectively, by using the appropriate 
stellar mass--to--light ratio, $\Upsilon_{K}$. The application of population 
synthesis techniques to disk galaxy evolution models shows 
that $\Upsilon_{K}$ depends mainly on the integral color (Tinsley 1981; 
Bruzual 1983; Bell \& de Jong 2001). In ZAHF we have used a
$\Upsilon_{K}$ inferred from the latter paper. In a more recent
work, Bell et al. (2003a) used galaxy evolution model fits for a large sample 
of galaxies from the Two Micron All Sky Survey (2MASS) and the Sloan Digital Sky Survey
(SDSS). They obtained a correlation of $\Upsilon_{K}$ with $(B-R)$ color 
which is shallower and more scattered than that in Bell \& de Jong (2001). Unfortunately,
Bell et al. (2003a)  did not report the correlation of $\Upsilon_{K}$ 
with the $(B-K)$ color. We have obtained this correlation by a re--scaling
procedure that makes use of the $\Upsilon_{K}$-$(B-R)$ correlation given in Bell 
at al. (2003a), and the $\Upsilon_{K}$-$(B-R)$ and $\Upsilon_{K}$-$(B-K)$ correlations 
given in Bell \& de Jong (2001; their model ``formation epoch with bursts'' was used).  
We obtain the following result: 
\begin{equation} \label{gammaHSB}
{\rm Log}\Upsilon_{K}=-0.38 + 0.08(B-K).
\label{gammaK}
\end{equation}

For low SB values, at least for blue galaxies, $\Upsilon_{K}$ seems to 
anti-correlate
with color (Verheijen 1997). In Fig. 1 of Bell \& de Jong (2001), it is
also shown how the inferred values of $\Upsilon_{K}$, for low SBs, no longer
decrease with decreasing SB, and even start to increase
as SB decreases, although the scatter is large.  In the absence of
detailed models, we approximate the $\Upsilon_{K}$-$(B-K)$ correlation for
LSB galaxies in such a way that for $(B-K)> 3$, $\Upsilon_{K}$ is given by 
eq. (\ref{gammaHSB}), while for $(B-K)\le 3$,
\begin{equation} \label{gammaLSB}
\Upsilon_{K}=1.90-0.40(B-K).
\end{equation}
The latter dependence is from a linear eye--fit to the $\Upsilon_{K}$-$(B-K)$ 
correlation of the {\it blue} LSB galaxies in Verheijen (1997), 
for his case of a Hernquist halo model and the constrained decomposition 
method. After using the new $\Upsilon_{K}$-$(B-K)$ correlations, the stellar
and baryonic masses of the galaxy calculated here are slightly different from those 
used in ZAHF.  
The scale length of the stellar disk, \hs, is assumed to be equal 
to the scale length in the band $K$, \hK. Most of the stars in the disk 
are properly traced using this band.

The disk gas mass, \mg, is estimated as:
\begin{equation} \label{mg}
\mg = 1.4 M_{\rm HI}\left[ 1+ \frac{M_{\rm H_2}}{M_{\rm HI}} \right]
\end{equation}
where the factor 1.4 takes into account helium and metals, and M$_{\rm H_2}$ is
the mass in molecular hydrogen. The M$_{H_2}/$M$_{HI}$
ratio has been found to depend on the morphological type $T$
(Young \& Knesek 1989). Using the latter paper, McGaugh \& de
Blok (1997) estimated that M$_{\rm H_2}/$M$_{\rm HI} = 3.7 -0.8T + 0.043T^{2}$.
For T$ < 2$, this empirical fitting formula could overestimate
the gas mass in galaxies, thus we assume: M$_{\rm H_2}/$M$_{\rm HI} = 2.3$ (the 
value of the fit at $T=2$) for $T< 2$.

The galaxy baryonic mass is defined as \md\ = \ms\ + \mg, and the
gas mass fraction as \fg\ = \mg/\md. Unfortunately, we do not
have information on the gas surface density profile parameters 
($\Sigma_{g,0}$ and h$_g$)
for the galaxies in our sample.  We need them in order to calculate the 
baryonic disk central surface densities, \sigb, and scale lengths, \hb. 
As discussed in
ZAHF, we assume that the total gas surface density follows an
exponential distribution with a scale length 3 times that of \hK. Thus,
$\Sigma_{\rm g,0}= \mg/2\pi(3\hK)^2$. The baryonic quantity \sigb~ is then calculated 
as \sigb\ = \sigs\ + $\Sigma_{\rm g,0}$. The
corresponding baryonic radius will be then
$\hb = \hK[(\sigs + 9\Sigma_{\rm g,0})/\sigb)]^{0.5}$. In fact, the addition
of the (uncertain) gas disk parameters will have little impact on our final
results. This inclusion is important only for gas rich galaxies (late--type,
blue LSB galaxies). We believe that in spite of the uncertainties, some real 
systematical variations of the baryonic disk scale length 
and surface density with the gas content of galaxies can still be taken
into account through our procedure. 

We have also estimated the errors in the composite quantities
by adequately propagating the errors of the primary quantities. 
See Appendix A for details.

\section{The scaling relations and their dispersions}

For the sample of 76 galaxies presented in \S 2, we proceed to
construct the physical scaling relations $\vm$-$L$ (-$M$), $R$-$L$ (-$M$), and 
$\vm$-$R$ in the bands $B$ and $K$, and for stars and baryons. 
Some authors recommend to use {\bf $V_{\rm flat}$}, the outer asymptotic flat part of the 
rotation curve, as an estimate of the circular velocity, instead of \vm. We stress 
that for studies aiming to understand the origin of the TFR and its scatter (the same 
applies for the $\vm$-$R$ relation) as well as to compare them with 
galaxy model predictions, the adequate quantity to be used is \vm\ {\it because 
it maximizes the contribution of the disk component for a given halo}.
By using $V_{\rm flat}$, which traces the dynamics of the galaxy in outer regions, 
where the halo component tends to dominate, we lose information about the disk contribution. 
In fact, here we use neither \vm\ nor $V_{\rm flat}$, but \W. However, as 
mentioned in \S 2, there is some evidence that $\W\approx 2\vm$.  

For the luminosities
and masses, here we use the total ones, i.e. the sum of disk and bulge.
For the cases when bulges form by disk secular evolution processes (e.g. Avila-Reese
\& Firmani 2000; Avila-Reese et al. 2005),  this is 
a good approximation for total disk luminosities and masses. 
For the scale length, the corresponding $B$ and $K$ band radii are used
in the luminous scaling relations. For the stellar and baryonic scaling 
relations we assume respectively \hs=\hK\ and a scale length, \hb, which takes 
into account the sum of the stellar and gas disks (see \S 2.2).  

\begin{figure*}
\vspace{15.3cm}
\hspace{14.5cm}
\includegraphics{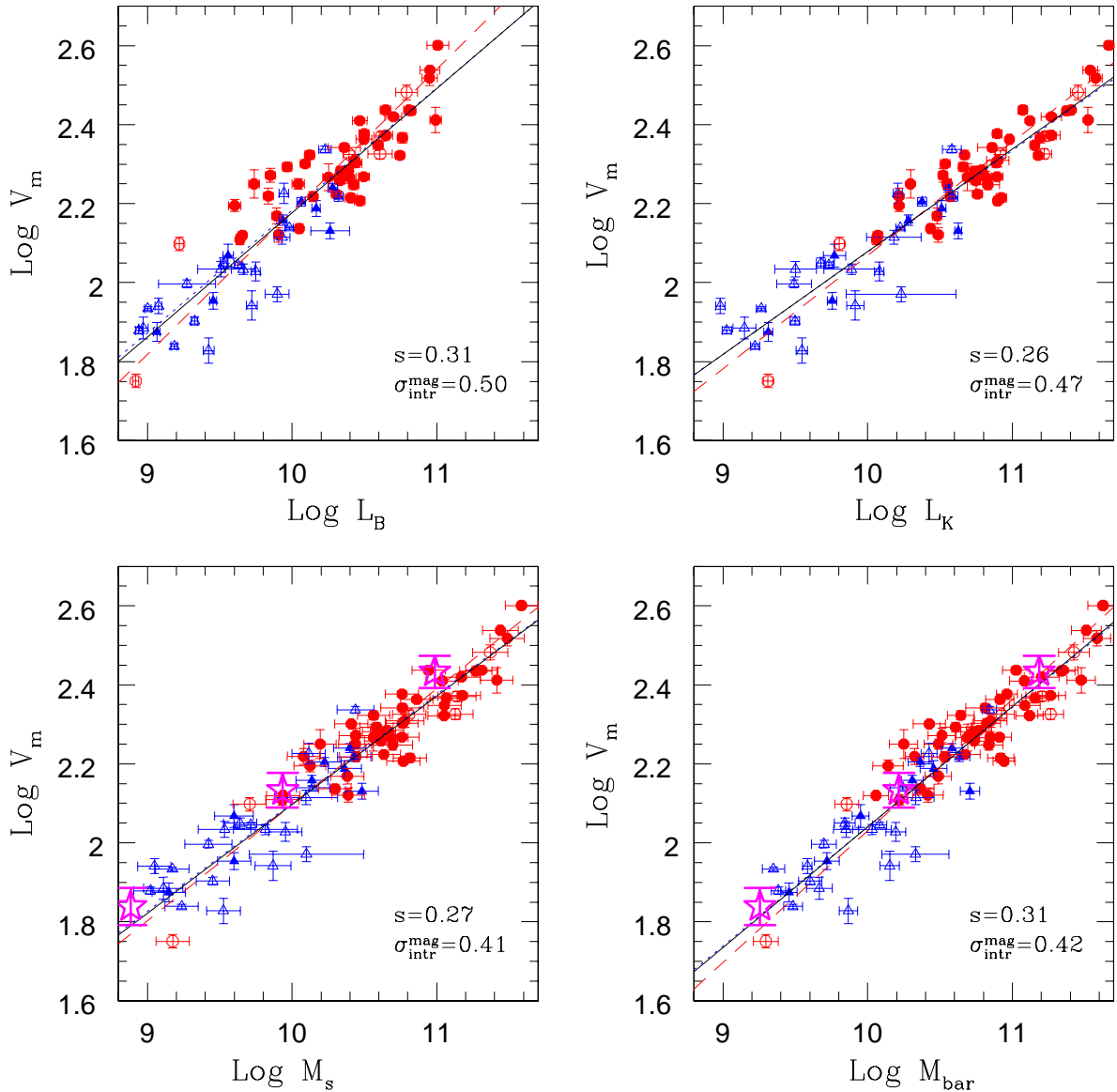}
\figcaption{Observed TF correlations in the $B$ and $K$ bands (upper panels) 
and for stellar and baryonic masses (lower panels). \vm~ is in km/s and the luminosities
and masses in the corresponding solar units. Solid and empty symbols are for 
HSB ($\muK\le 18.5$ mag/arcsec$^2$) and LSB ($\muK> 18.5$ mag/arcsec$^2$ ) galaxies,
respectively. Red galaxies ($[B-K]>3$) are represented with (red) circles, while blue galaxies
($[B-K]\le 3$) with (blue) triangles.  The solid, dotted, and dashed lines are the corresponding
orthogonal, forward, and inverse linear heteroscedastic doubly--weighted regressions, 
respectively. The slopes and inverse intrinsic scatters of the orthogonal fits are given
inside each panel. The solid (magenta) stars with vertical error bars
in the lower panels are the average values of \vm~ and the $rms$ scatter for a given mass
from the \LCDM--based disk galaxy evolutionary models presented in FA-R  ($\sigma_8=1$). The 
models were rescaled to $h=0.7$.   
\label{TF}
}
\end{figure*}

\begin{figure*}
\vspace{15.3cm}
\hspace{14.5cm}
\includegraphics{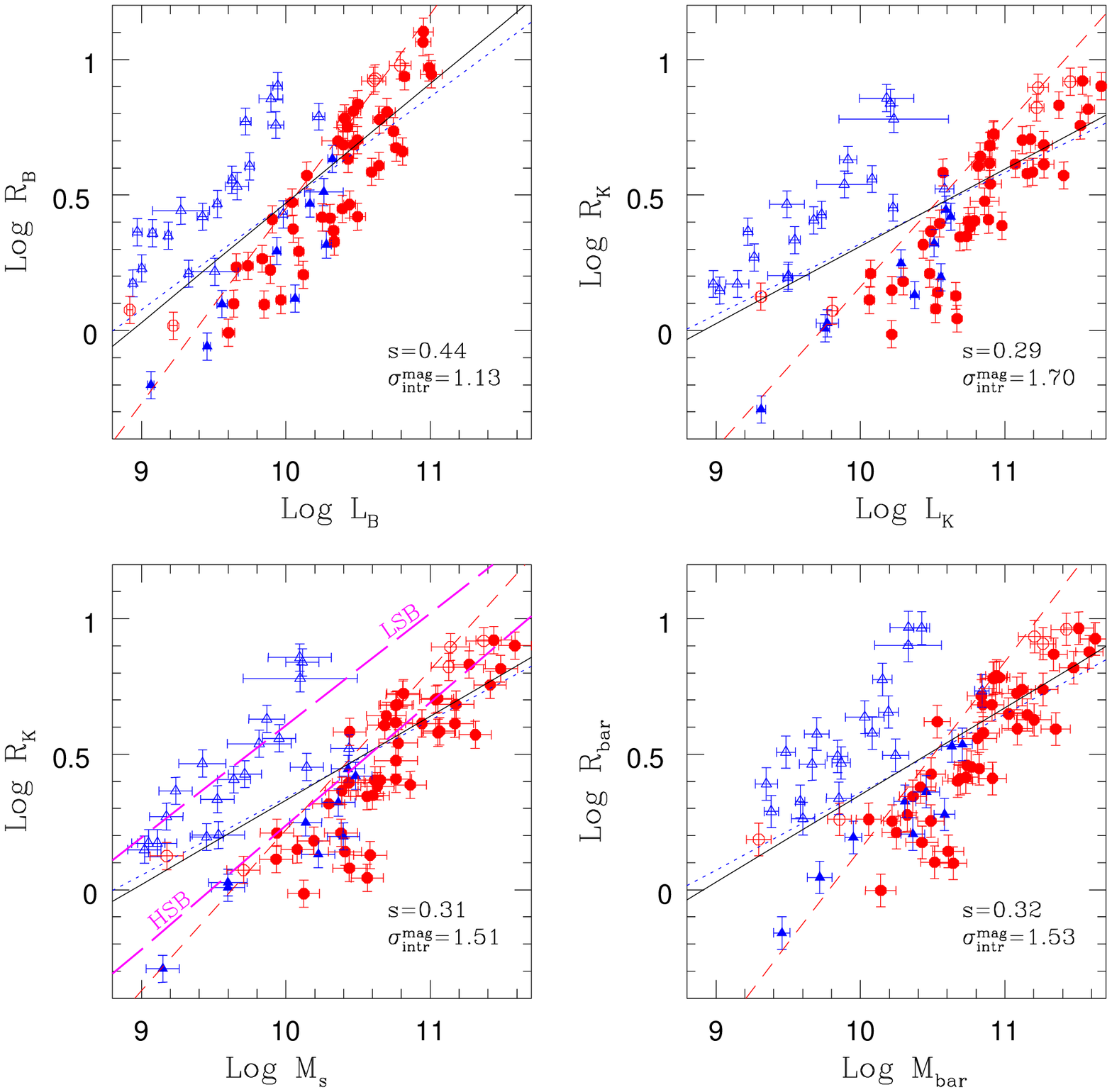}
\figcaption{
Observed radius--luminosity correlations in the $B$ and $K$ bands (upper panels), 
and radius--mass correlations for stellar and baryonic quantities (lower panels). 
Symbols, lines types, labels, and units are as in Fig. \ref{TF}. Radii are in kpc.
The thick dashed magenta lines in the
\hK-\ms~ diagram are the fits to the model LSB and HSB+very HSB galaxies 
presented in FA-R (rescaled to $h=0.7$).
\label{RL}
}
\end{figure*}

In Figs. \ref{TF}--\ref{VR}, the observational points with the 
estimated error bars are plotted in the diagrams corresponding to the disk galaxy 
scaling relations studied here. The variables are given logarithmically. 
Circles and triangles are used for galaxies with $(B-K) > 3.0$ (red) and $\le 3.0$ (blue), 
respectively. The open and solid symbols are for galaxies with 
$\muK\le 18.5$ (HSB) and $\muK> 18.5$ (LSB), respectively. 
To quantify the correlations and compare them among the different
pass--bands and mass quantities (and eventually to model predictions), 
one needs to fit the correlations. To this end, a linear regression analysis should
be used. We recall that our aim here is to explore variations for the scaling 
relations in the different cases ($B$ and $K$ bands, stars, and baryons) rather 
than study in detail the relations for each one of the cases. Therefore, it is 
important for us {\it to use the same regression method for all the cases rather than 
to choose an optimal method}.

For studies aimed at theoretical interpretations it is better to use a linear 
fit that is symmetric to interchanges of the two variables, e.g. the bisector
or orthogonal regression linear methods. Here we fit the
correlations using the forward, backward, bisector, and orthogonal linear
regression models. However, for further analysis and discussion, {\it we will use 
only the orthogonal regression results}. The origin of coordinates for each
correlation is shifted to the 'barycenter' of the corresponding variables, 
where the slope and zero point are uncorrelated.

The corresponding linear regressions, {\it not taking into account} the errors in 
the variables, were carried out for our galaxy sample in Zavala (2003) 
by using the SLOPES routine (Isobe et al. 1990; Feigelson \& Babu 1992). This 
seems to be a feasible approximation when the intrinsic scatter in the 
regression line dominates any errors arising from the measurement and error 
propagation procedures (Isobe et al. 1990). The TFR is the 
tightest one among the scaling relations. Studies of the TFR have 
shown indeed that the intrinsic scatter of the linear fit is larger than
the measurement and correction errors (for both magnitude and velocity).
For example, Giovanelli et al. (1997) showed that the latter errors account
for $\sim 0.25-0.15$ mag (from smaller to larger velocities), while the 
total scatter to the fit is $\sim 0.45-0.30$ mag for the same range
of velocities, implying that the {\it intrinsic} scatter dominates over
the measurement and correction uncertainties. For the other scaling
relations, this difference is expected to be even larger.  
 
Since we have obtained estimates of the errors in the variables of the scaling relations 
(Appendix A), here we carry out the corresponding ``error--in--variable'' 
regression models. Thus, the analysis in the present paper is more exact than 
in Zavala (2003) and it allows us to roughly estimate the values 
of the intrinsic scatter in the different scaling relationships. 
Because the errors differ from point to point, the model should be heteroscedastic. 
We assume that the scaling relations have an (unknown) intrinsic 
scatter, which we want to estimate, explore, and confront with theory. 
The fitting regressions taking into account heteroscedastic measurement 
errors are carried out using the method and routine presented in Akritas \& 
Bershady (1996). The total average intrinsic scatter, \sigintr, is estimated as
the square root of the variance of the fit, $\sigma_{fit}^2$, subtracted from the average 
(in both variables) observable variance, $\sigma_{obs}^2$ 
($=N/\sum 1/[\sigma_y^2 + b^2\sigma_x^2]$, where $N$ is the number of data points and $b$ 
is the slope of the correlation), i.e., $\sigma_{intr}^2= \sigma_{fit}^2 - \sigma_{obs}^2$. 
More elaborate fitting procedures could be used to estimate the intrinsic scatter
(e.g., Gnedin et al. 2007).
However, given the level of approximation we have used to determine the errors for the
data, we consider that our first--order approximation is enough. 

Tables $1-3$ show the results from the different regression methods applied
to the three scaling correlations in the four cases (baryonic and stellar
quantities, and $K$ and $B$ bands). 
We report the zero point, $a$, the slope and $b$, with their respective 
standard deviations (calculated in the 'barycenter' of the variables), as well as 
the corresponding square roots of the variance of the fit, 
\sigfit, the bi--weighted average of the individual (in both variables) variances, 
\sigobs, and the total average intrinsic variance, \sigintr.
We recall that our analysis, including the scatters, refers to the logarithm of the 
quantities involved.  In the
panels of Figs. \ref{TF} to \ref{VR}, we plot the forward (dashed line),
backward (dotted line), and orthogonal (solid line) doubly--weighted linear fits.

\begin{deluxetable}{llllllll}
\tablewidth{8.3cm}
\tablecaption{Linear doubly--weighted fit parameters and scatters for the 
TF (Log--Log) correlations.}
\label{tab:TF}
\tablehead{\colhead{Fit} & \colhead{$b$} & \colhead{$\pm 1\sigma$} &
\colhead{$a$} & \colhead{$\pm 1\sigma$} & \colhead{\sigobs} & \colhead{\sigfit} 
& \colhead{\sigintr} } 
\startdata
{\vm-\md} \\ 
  Forward  & 0.303 & 0.012 & -0.979 & 0.127 & 0.029 & 0.058  & 0.050 \\
  Inverse  & 0.333 & 0.014 & -1.300 & 0.151 & 0.031 & 0.063  & 0.055 \\
  Bisector & 0.317 & 0.012 & -1.140 & 0.132 & 0.030 & 0.060  & 0.052 \\
  Orthog.  & 0.306 & 0.012 & -1.024 & 0.129 & 0.028 & 0.058  & 0.051 \\
{\vm-\ms}\\
  Forward  & 0.274 & 0.011 & -0.639 & 0.121 & 0.036 & 0.058  & 0.045 \\
  Inverse  & 0.294 & 0.013 & -0.843 & 0.137 & 0.038 & 0.061  & 0.048 \\
  Bisector & 0.284 & 0.012 & -0.741 & 0.127 & 0.037 & 0.059  & 0.046 \\
  Orthog.  & 0.274 & 0.012 & -0.650 & 0.123 & 0.035 & 0.058  & 0.045 \\  
{\vm-$L_K$} \\
  Forward  & 0.258 & 0.012 & -0.503 & 0.126 & 0.014 & 0.052  & 0.050 \\
  Inverse  & 0.272 & 0.012 & -0.652 & 0.132 & 0.015 & 0.057  & 0.054 \\
  Bisector & 0.272 & 0.012 & -0.652 & 0.132 & 0.014 & 0.054  & 0.052 \\
  Orthog.  & 0.261 & 0.012 & -0.520 & 0.128 & 0.014 & 0.051  & 0.049 \\  
{\vm-$L_B$} \\
  Forward  & 0.310 & 0.015 & -0.917 & 0.152 & 0.016 & 0.064  & 0.062 \\
  Inverse  & 0.361 & 0.016 & -1.430 & 0.169 & 0.018 & 0.073  & 0.072 \\
  Bisector & 0.335 & 0.014 & -1.170 & 0.149 & 0.017 & 0.068  & 0.066 \\
  Orthog.  & 0.314 & 0.015 & -0.963 & 0.154 & 0.016 & 0.065  & 0.063 \\ 
\enddata
\tablecomments{Log\vm$=a+b$Log$M$($L$). \vm~ is given in \kms, \md~ and \ms~ in 
\msun, $L_K$ in L$_{K\odot}$, and $L_B$ in L$_{B\odot}$.}
\end{deluxetable}

\begin{deluxetable}{llllllll}
\tablewidth{8.3cm}
\tablecaption{Linear doubly--weighted parameters and scatters for the radius--mass (luminosity)
(Log-Log) relations}
\label{tab:RL}
\tablehead{\colhead{Fit} & \colhead{$b$} & \colhead{$\pm 1\sigma$} &
\colhead{$a$} & \colhead{$\pm 1\sigma$} & \colhead{\sigobs} & \colhead{\sigfit} 
& \colhead{\sigintr}} 
\startdata
{\hb-\md} \\ 
  Forward  & 0.287 & 0.035 & -2.510 & 0.379 & 0.065 & 0.207  & 0.196 \\
  Inverse  & 0.692 & 0.098 & -6.770 & 1.030 & 0.085 & 0.032  & 0.313 \\
  Bisector & 0.474 & 0.042 & -4.470 & 0.444 & 0.073 & 0.236  & 0.225 \\
  Orthog.  & 0.322 & 0.039 & -2.870 & 0.422 & 0.067 & 0.208  & 0.197 \\
\hs-\ms \\
  Forward  & 0.285 & 0.031 & -2.510 & 0.327 & 0.061 & 0.196  & 0.187 \\
  Inverse  & 0.592 & 0.073 & -5.700 & 0.765 & 0.087 & 0.282  & 0.269 \\
  Bisector & 0.430 & 0.034 & -4.020 & 0.366 & 0.072 & 0.218  & 0.206 \\
  Orthog.  & 0.310 & 0.033 & -2.770 & 0.352 & 0.063 & 0.196  & 0.187 \\  
\hK-$L_K$  \\
  Forward  & 0.262 & 0.030 & -2.300 & 0.324 & 0.052 & 0.201  & 0.194 \\
  Inverse  & 0.593 & 0.075 & -5.770 & 0.794 & 0.056 & 0.294  & 0.289 \\
  Bisector & 0.418 & 0.035 & -3.940 & 0.378 & 0.054 & 0.223  & 0.217 \\
  Orthog.  & 0.285 & 0.033 & -2.540 & 0.355 & 0.052 & 0.201  & 0.194 \\  
\hB-$L_B$ \\
  Forward  & 0.393 & 0.039 & -3.460 & 0.403 & 0.053 & 0.206  & 0.198 \\
  Inverse  & 0.718 & 0.065 & -6.730 & 0.658 & 0.058 & 0.278  & 0.271 \\
  Bisector & 0.545 & 0.037 & -4.980 & 0.387 & 0.055 & 0.223  & 0.216 \\
  Orthog.  & 0.441 & 0.044 & -3.940 & 0.452 & 0.054 & 0.207  & 0.200 \\ 
\enddata
\tablecomments{Log$R=a+b$Log$M$($L$). The radii are given in kpc, \md~ and \ms~ 
in \msun, $L_K$ in L$_{K\odot}$, and $L_B$ in L$_{B\odot}$.}
\end{deluxetable}

\begin{deluxetable}{llllllll}
\tablewidth{8.3cm}
\tablecaption{Linear doubly--weighted regression parameters and scatters for the \vm-radius 
(Log-Log) relations}
\label{tab:VR}
\tablehead{\colhead{Fit} & \colhead{$b$} & \colhead{$\pm 1\sigma$} &
\colhead{$a$} & \colhead{$\pm 1\sigma$} & \colhead{\sigobs} & \colhead{\sigfit} 
& \colhead{\sigintr}}
\startdata
\vm-\hb \\ 
  Forward  & 0.332 & 0.078 & 2.030 & 0.044 & 0.024 & 0.160  & 0.158 \\
  Inverse  & 1.520 & 0.318 & 1.420 & 0.174 & 0.093 & 0.359  & 0.348 \\
  Bisector & 0.768 & 0.056 & 1.810 & 0.044 & 0.048 & 0.201  & 0.195 \\
  Orthog.  & 0.506 & 0.109 & 1.937 & 0.063 & 0.033 & 0.167  & 0.164 \\
\vm-\hK \\
  Forward  & 0.390 & 0.068 & 2.030 & 0.035 & 0.024 & 0.149  & 0.147  \\
  Inverse  & 1.220 & 0.196 & 1.670 & 0.096 & 0.063 & 0.274  & 0.267  \\
  Bisector & 0.727 & 0.053 & 1.880 & 0.037 & 0.039 & 0.177  & 0.172  \\
  Orthog.  & 0.534 & 0.085 & 1.970 & 0.046 & 0.030 & 0.155  & 0.152  \\  
\vm-\hB \\
  Forward  & 0.378 & 0.057 & 2.010 & 0.035 & 0.023 & 0.142  & 0.140  \\
  Inverse  & 1.050 & 0.157 & 1.680 & 0.088 & 0.055 & 0.257  & 0.241  \\
  Bisector & 0.662 & 0.049 & 1.870 & 0.038 & 0.036 & 0.167  & 0.163  \\
  Orthog.  & 0.481 & 0.067 & 1.960 & 0.043 & 0.028 & 0.146  & 0.143  \\  
\enddata
\tablecomments{Log\vm$=a+b$Log$R$.  \vm~ is given in \kms, and the radii in kpc.}
\end{deluxetable}

\subsection{Luminosity (mass) vs velocity (TFRs)}

As Fig. \ref{TF} and Table 1 show, the \vm-$L_B$, \vm-$L_K$, \vm-\ms~ and 
\vm-\md~ correlations are tight, although the differences between the lines 
obtained from the different regression methods are slightly larger than the
dispersion associated with any given line. Therefore, the choice of the fitting method will 
slightly affect the results. For all the cases, the slopes of the forward
correlations are shallower than those of the inverse correlations. For the bisector
and orthogonal correlations, their slopes are between those of the forward
and inverse correlations. The orthogonal correlations have shallower slopes
than the bisector correlations, which are very close to those obtained from the
forward fit. In the following, we analyze only the correlations
fitted with the orthogonal regression method. A deep statistical analysis of each one
of the correlations presented here is out of the scope of the paper. 

From the baryonic \vm-\md~ to stellar \vm-$\ms$ TFR, the slope 
is shallower by  $\approx 11\%$, and the mass at \vm = 160 km/s is 1.36 times (0.134 dex)
smaller; for larger velocities, the mass difference becomes smaller. 
From \vm-\ms~ to \vm-$L_K$, the slope remains almost constant. From  
\vm-$L_K$ to \vm-$L_B$ the slope is steeper by $\approx 20\%$, and some 
segregation by color appears. 

Keeping in mind that most of the 
previous works used to report the forward slopes of the $L$-($M$-)\vm~ correlations,
we remark that the equivalent slopes here would be the inverse of our inverse
linear regressions (see Table 1): 3.00, 3.40, 3.67, and 2.77 for the baryonic, stellar, $K-$ 
and $B-$band cases, respectively.

It should be noted that the sample used here is broad in galaxy properties
and {\it it was not pruned to reduce the TFR scatter}. The few works that 
carry out analysis of samples similar
to ours, although with different fitting methods, give slope values for the 
luminous TFRs (Kannappan, Fabricant \& Franx 2002; Verheijen \& Sancisi 2001; Dutton et 
al. 2007; Courteau et al. 2007) and for the stellar TFR 
(Pizagno et al. 2005; Gnedin et al. 2007) that are close to our estimates. The recent work 
by Pizagno et al. (2007), that analyzes a sample of 162 SDSS disk galaxies with H$\alpha$
rotation curves, reports slopes of the luminous TFRs that are much steeper than those 
obtained here. 
The baryonic TFR obtained in McGaugh (2005) is shallower than ours, however, he uses 
$V_{\rm flat}$ instead of \vm.

The estimated average intrinsic scatters, \sigintrV, in the \vm-\md~, 
\vm-$\ms$, \vm-$L_K$, and \vm-$L_B$ correlations are 0.051, 0.045, 0.049 and 
0.063 dex, respectively. The intrinsic scatter is larger (by $\sim 10\%$)
for the baryonic correlation than for the stellar one; the largest scatter is in 
the $B-$band. The intrinsic scatter along the $M$ or $L$ axis expressed in magnitudes, 
$\sigma^{\rm mag}_{\rm intr}=2.5\times \sigintrV/b$, oscillates between 0.41 mag 
(stellar correlation) and 0.5 mag ($B$ band correlation) for all 
the regression methods used here (see values for the orthogonal fit in Fig. \ref{TF}).   

Recall that our observational error estimates and their propagation are
approximations. Nonetheless, our estimates for the intrinsic scatter agree with other 
similar works. For example, Dutton et al. (2007) estimate $\sigintrV\approx 0.052$ in 
the $I-$band, which falls in between our values in the $B$ and $K$ bands. 
Pizagno et al. (2007), 
who carried out a deep analysis of the errors for their sample, find typical values of 
the intrinsic scatter for the optical/NIR TFRs of 0.40--0.45 mag, which is 
close to our findings. 

\subsection{Radius vs luminosity (mass)}

The \hB-$L_B$, \hK-$L_K$, \hs-$\ms$, and \hb-\md~ correlations are 
scattered and strongly segregated by disk central SB (Fig. \ref{RL}; 
Table 2). For a given luminosity or mass, the disk scale length, or central SB,
span a large range of values.  We see from Fig. \ref{RL} that a segregation by 
color is also present. There is no significant difference between 
the slopes and zero points of the baryonic and stellar $R$-$M$ correlations. 
Both the mass and the radius increase when changing from stellar to
baryon quantities, but the change is small and shifts galaxies along the same 
correlation, whose slope is $\approx 0.32$.  From the stellar to the $K-$band 
correlation, the slope slightly decreases, while from the $K-$ to the $B-$band 
correlation the slope strongly increases, from $0.28\pm 0.03$ to $0.44\pm 0.04$.

The intrinsic scatter measured in the Log of the radii is approximately
the same in all the cases, $\sigintrR\approx 0.2$.  When translated to
the ($X-$axis) scatter in Log$L$ or Log$M$ and expressed in magnitudes, the
scatter is around 1.1 mag for the $B-$band correlation and 1.5--1.7 mag for the
other correlations. Hence, in the $B$ band, the dispersion along the luminosity
is decreased by some compensation effect. The observational errors in the $R$-$L$(-$M$) 
diagrams are significantly smaller than the corresponding  average intrinsic scatters. 
Note that, contrary to the TFRs, for the $R$-$L$ (-$M$) correlations, the use of a 
broad range in SBs (from HSB to LSB galaxies) significantly spreads
the correlation and influences the slope and zero--point values. 
This can be clearly appreciated in Fig. \ref{RL}: if we exclude the
LSB galaxies, then the fit changes notably. In most of the previous works 
aimed to estimate the disk scaling relations, {\it LSB galaxies were not considered.
This is unfortunate if the observations are to be used to compare
and constrain theoretical models.}  Since the galaxy morphological type correlates 
with the SB, one expects that the $R$-$L$ correlation also will depend on the type;
Graham \& Worley (2008) have indeed shown this explicitely.

\begin{figure*}[htb!] 
\vspace{7.5cm}
\hspace{15.3cm}
\includegraphics{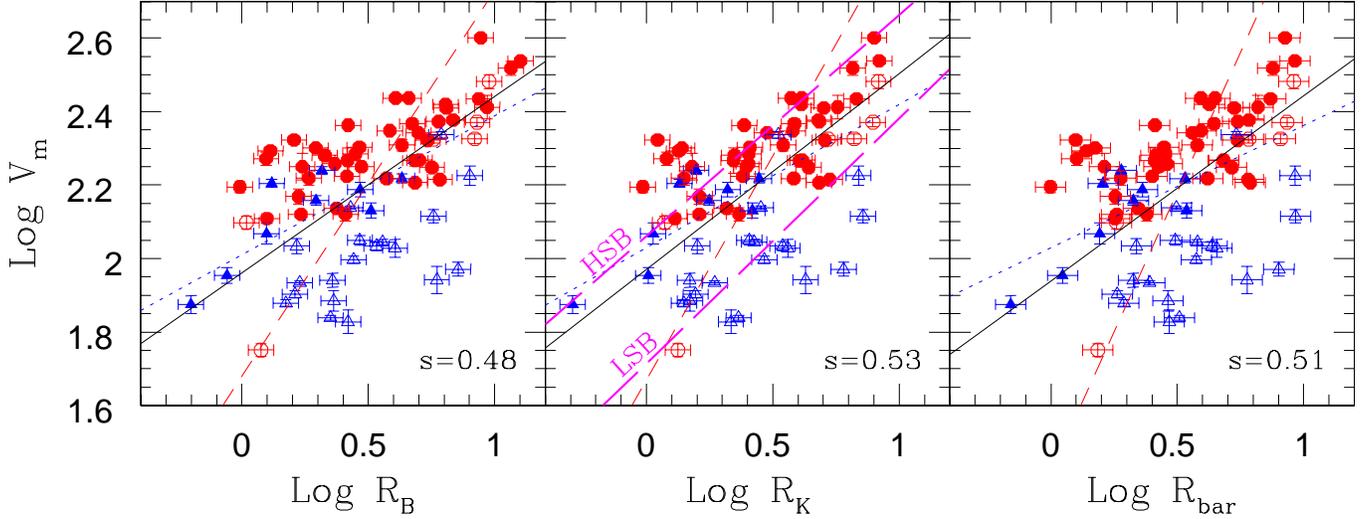}
\figcaption{Observed \vm--radius correlations in the $B$ and $K$ bands (left and mid panels), 
and baryonic \vm--\hb~ correlation (right panel). Because we assumed \hs=\hK, the 
stellar \vm-\hs~ correlation is as the \vm--\hK~ one. Symbols, line types and labels are 
as in Fig. \ref{TF}. The thick dashed magenta lines are
the fits to the model LSB and HSB+very HSB galaxies presented in FA-R (rescaled to $h=0.7$). 
\label{VR}
}
\end{figure*}

\subsection{Velocity vs radius}

The \vm-\hB, \vm-\hK, and \vm-\hb~ correlations are shown
in Fig. \ref{VR} and their fitting parameters are given in Table 3. 
The \vm-\hs~ correlation is the same as the \vm-\hK~ one because
we assumed that $\hs=\hK$. The \vm$-R$ correlations are highly
scattered, and segregated by disk surface brightness and color. 
For these correlations, the differences between forward and inverse 
fits are large.
The relative errors in the fitted slope and zero point parameters 
are the largest ones among the three scaling relations studied here.

Because of the large dispersions in the fitted slopes and zero points of
the \vm$-R$ correlations, 
there is no significant difference among these parameters 
when we change from the $B$ to the $K$ band and to the baryonic case.
For the orthogonal regression, the 
slopes and zero points are $\approx 0.5$ and $\approx 1.95$, respectively. 
The intrinsic scatter slightly decreases from the baryonic to the
$K-$band, \sigintrV$\approx 0.17$ and 0.15, respectively.
The estimated inverse intrinsic scatter (along the $L$ axis) increases
from the $K$ to the $B$ band.

\subsection{Other global correlations}

\begin{figure*}
\vspace{8.8cm}
\hspace{15.3cm}
\includegraphics{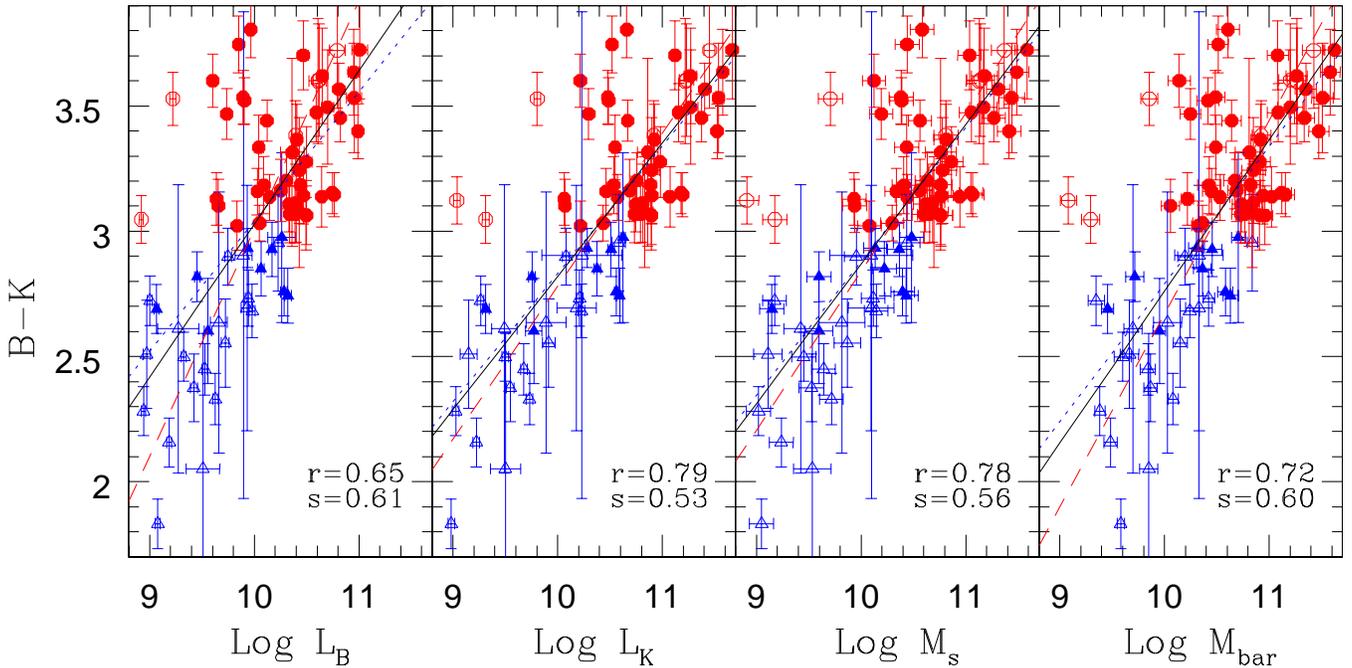}
\figcaption{Corrected $(B-K)$ color vs the logarithms of $L_B$, $L_K$, \ms, and \md.
Symbols and line types are as in Fig. \ref{TF}. The Pearson's correlation coefficient, 
$r$, and the slope of the orthogonal fit, $s$,  appear in the right-bottom corner of 
each panel.
\label{colorL}
}
\end{figure*}

In order to complement the presented scaling correlations and their changes when 
moving from luminous to stellar and baryonic quantities, we explored the possibility 
of other correlations for our galaxy sample involving luminosities (masses), global 
color, central surface densities (brightnesses), and stellar mass content. 
Fig. \ref{colorL} shows the $(B-K)$ color vs $L_B$, $L_K$, \ms, and \md, with 
the corresponding Pearson correlation coefficients and slopes. In spite of the 
$L$--dependent correction by internal extinction that we have applied, a clear 
correlation of color with $L$ and $M$ remains: more massive (luminous) galaxies are 
on average redder than the less massive (luminous) ones. 
The $B-$band correlation is the most scattered,
as expected. The slope becomes slightly steeper when changing
from \ms~ to \md~ mainly because the less massive, bluer galaxies have typically
larger gas mass fractions, shifting more to the right side in the Log$M$ axis 
than the massive galaxies. The slope also becomes steeper when changing from 
$L_K$ to $L_B$, in this case because less luminous galaxies being bluer on average, 
shift more to the right side in the Log$L$ axis than the redder luminous galaxies.

\begin{figure*}
\vspace{15.5cm}
\hspace{15.3cm}
\includegraphics{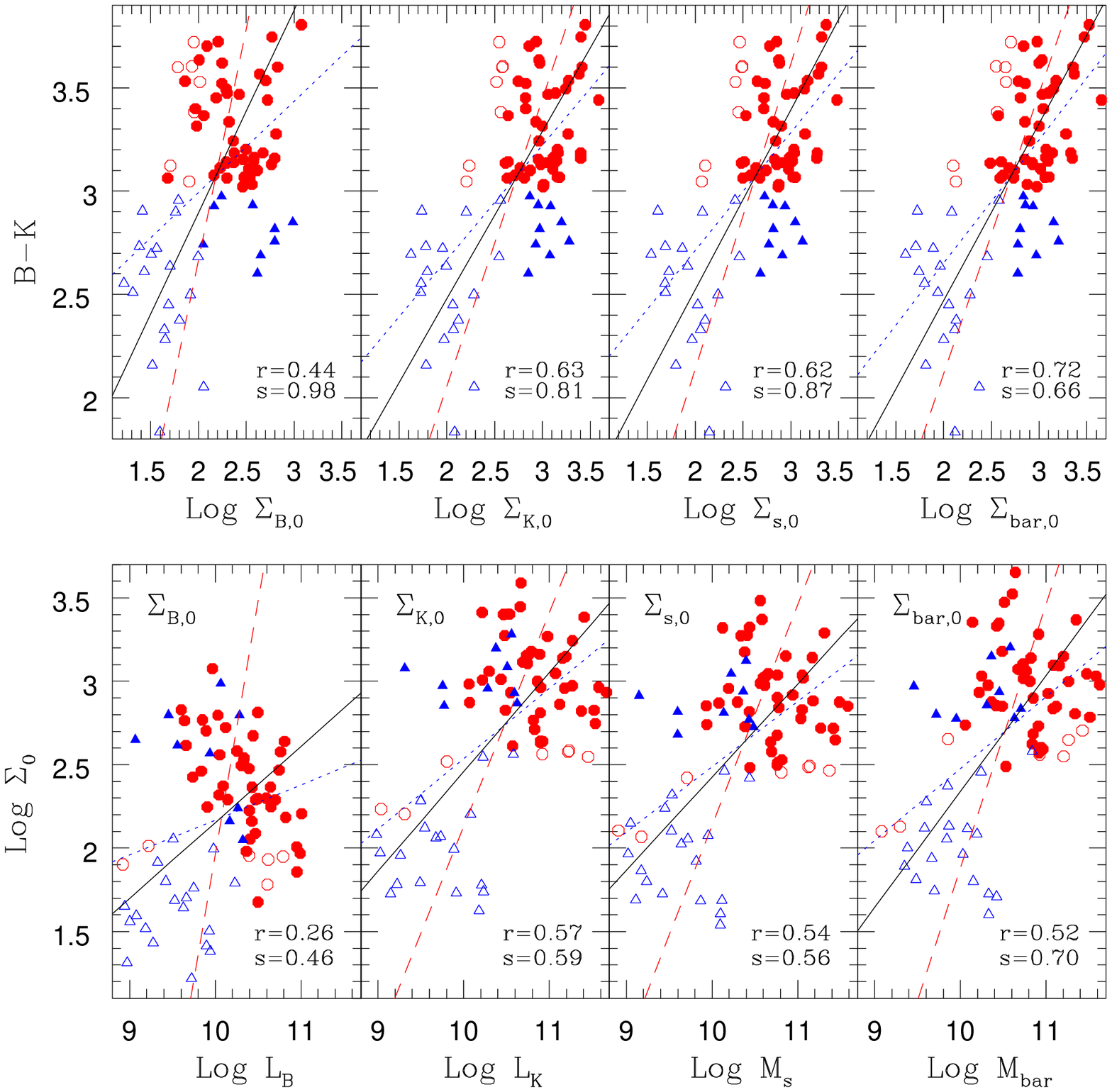}
\figcaption{Corrected $(B-K)$ color vs the Log of \sigB, \sigK, \sigs 
and \sigb~ (top panels), and Log$\Sigma_0$ vs Log$(M,L)$ for B, K bands, stellar 
and baryonic components from left to right (bottom panels). Symbols and line types are as in 
Fig. \ref{TF}. Labels are as in Fig. \ref{colorL}.
\label{colorSB}
}
\end{figure*}

Noisy Log--Log correlations are also observed between $(B-K)$ and the different 
SBs and surface densities, and between the latter quantities and the luminosities 
and masses (Fig. \ref{colorSB}). The slopes of the orthogonal fits are given inside
each corresponding panel. The general trend is that the higher the surface
density (or SB), the redder and more massive (luminous) is the galaxy. What lies on 
the basis of these trends for disk galaxies?  
We have also explored correlations related to the galaxy stellar's mass fraction,
\fs~ (or gas mass fraction $\fg=1-\fs$). This quantity is an important property of galaxies
related to their gas infall and SF rate histories; it will be discussed
in \S\S 4.2 (see Fig. \ref{stfrac} therein). The fraction \fs~ correlates strongly
with the $(B-K)$ color and weakly with the surface densities (or SBs)
and luminosities (masses), in that order. For all cases, the lowest SB galaxies 
break the correlations, suggesting a kind of a threshold for the gas surface density, 
below which the SF efficiency is almost independent of it (and of $M$ or $L$).


\subsection{Residual correlations and the search for other
significant parameters in the scaling relations}

To gain a more quantitative description of the disk galaxy scaling relations 
and their implications, we may explore the behavior of their residuals.
The possible correlations of the residuals among them, and other galaxy 
properties, bring valuable information on galaxy formation and evolution processes. 
In particular, the changes of the residual correlations when moving from optical 
to NIR bands and to stellar and baryonic quantities provide clues on the SF 
and stellar population evolution processes as well as on the dynamics 
of disk galaxies.

\begin{figure*}
\vspace{17.9cm}
\hspace{15.3cm}
\includegraphics{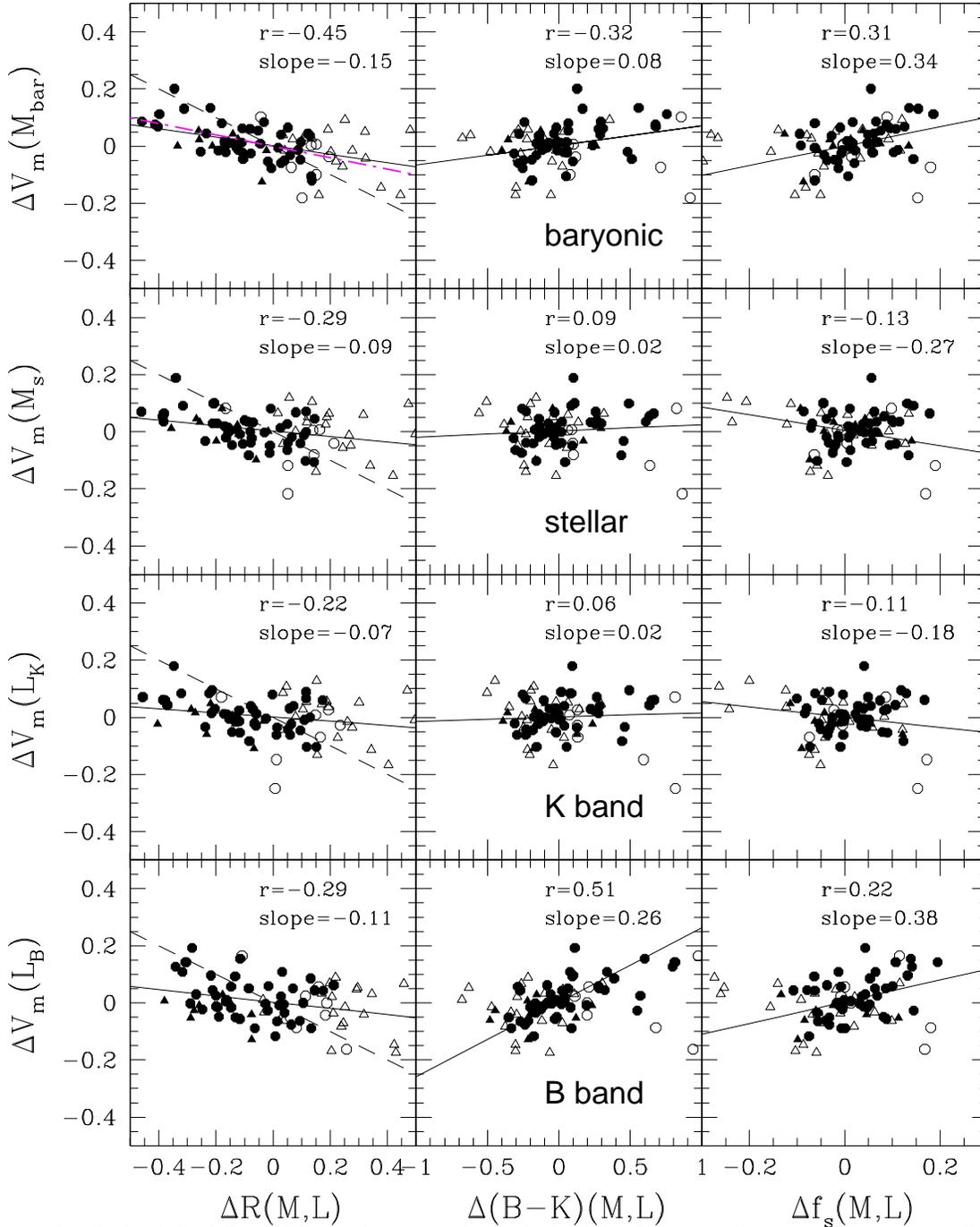}
 \figcaption{Residuals of the different TF correlations vs the residuals of the 
different radius--mass (luminosity), color--mass (luminosity), and 
\fs--mass (luminosity) correlations. From top to bottom, the rows correspond
to the baryonic, stellar, $K$ and $B$ band cases.  
Symbols are as in Fig. \ref{TF}. The solid lines are the orthogonal 
linear regressions. The corresponding Pearson correlation coefficients and the
slopes are given inside each panel. The dot--dashed (magenta) line in the upper left
panel is the linear fit to the model results by FA-R. The dashed lines in
all the panels in the left column indicate the prediction for an exponential
disk without dark halo (slope $-0.5$). 
\label{residuos}
}
\end{figure*}

In the different panels of Fig. \ref{residuos}, 
the residuals of the \vm-$M$(-$L$) correlations are plotted 
against the corresponding residuals of the $R$-$M$(-$L$), $(B-K)$-$M$(-$L$),
and $\fs$-$M$(-$L$) correlations for the baryonic, stellar, $K-$ and $B-$band 
cases, respectively. The orthogonal regression fit was used to obtain 
the residuals. As can be seen in Fig. \ref{residuos}, for most of the cases, the 
residuals are weakly correlated or completely uncorrelated with each other.

For the baryonic case, we find an anti--correlation of $\Delta$\vm(\md) with
$\Delta$\hb(\md) with a slope of $-0.15\pm 0.04$
(orthogonal linear regression): on average, the more a galaxy deviates in the 
direction of the high--velocity side in the \vm-\md~ relation, then the 
more it deviates in the direction of the small--radius 
side in the \hb-\md~ relation. It is illustrative to know
that for the extreme case of exponential disks without dark matter halos, the slope
would be $-0.5$, while for completely halo dominated galaxies, no correlation is
expected.  For the stellar and $K-$ and $B-$band cases, the trend described above 
almost disappears. These results show that {\it the residuals
of the \vm-$L$ (-$M$) and $R$-$L$ (-$M$) correlations can or cannot correlate, depending
not only on the inner dynamics of the galaxies themselves, but also on their SF 
histories, a question highlighted in FA-R}. From the left column of Fig. \ref{residuos},
we also learn that for all cases, the LSB galaxies 
are more scattered and uncorrelated than the HSB ones in the residuals plane.

As expected, the residuals of the $R$-$L$ (-$M$) correlations anti--correlate strongly 
with the corresponding disk central surface densities (or SBs) as:
$\Delta\hb(\md)\propto\Sigma_{\rm bar,0}^{-0.34}$, $\Delta\hs(\ms)\propto\Sigma_{\rm s,0}^{-0.39}$,
$\Delta\hK(L_K)\propto\Sigma_{\rm K,0}^{-0.37}$, and $\Delta\hB(L_B)\propto\Sigma_{\rm B,0}^{-0.45}$.
The surface density (or SB) is clearly a significant third parameter in all the
$R$-$L$ (-$M$) correlations. The tightest anti--correlation is for the baryonic 
case; then its strength decreases successively for the 
stellar, $K-$band, and $B-$band cases. The residuals of the
\vm-$R$ correlations correlate significantly with the corresponding surface densities
(or SBs). These residuals also correlate with those of the \vm-$L$ (-$M$)
correlations, but with a considerable segregation by SB (or surface density), 
resembling the $R$-$L$ (-$M$) correlations.

The findings reported above imply that the radius could be a statistically significant 
third parameter in the baryonic TFR. Applying a backward step--wise multiple linear 
regression procedure, where $F$ tests are used to compute the significance of each 
independent variable\footnote{
In this analysis, \vm~ is assumed to be the dependent variable and $M$ (or $L$),
$\Sigma_0$, $R$, $(B-K)$, $f_g$, and the morphological type $T$ are the independent
variables. For each case (baryonic, stellar, and $B$ and $K$ bands), one starts from 
a multiple variable linear regression using {\it all} the variables and then, through
the backward step--wise procedure, the variables that are not statistically
significant are eliminated step by step.}, Zavala (2003) has indeed found that 
the radius is a third (statistically significant) parameter in the 
baryonic TFR. However, the radius was no longer found to be a third parameter 
in the stellar and luminous TFRs. We confirm here the results by Zavala (2003), 
with the exception that for our analysis in the $B$ band, the radius is 
statistically more significant than in the analysis of Zavala (2003), however, 
we still find that the radius is not a third parameter in the $B-$band TFR; instead, 
the $(B-K)$ color is a third parameter in this case.

Since we are interested in the changes in the scaling correlations when changing
from one band to another and to stellar and baryonic quantities, it is important
then to analyze also the possible dependences of their residuals on galaxy
color and stellar fraction.  In the mid and right panels of Fig. \ref{residuos}, 
the residuals of the \vm-$L$ (-$M$) correlations are plotted vs the residuals of the 
$(B-K)$-$L$ (-$M$) and \fs-$L$ (-$M$) correlations, rather than just vs $(B-K)$
and \fs. In this way, the comparisons are at a given $L$ or $M$. In the
baryonic case, there are noisy trends: the residuals of the \vm-\md~ correlation 
increase on average for increasing  residuals of the $(B-K)$-\md~ and 
\fs-\md~ correlations (the galaxies become redder and with higher \fs); however, this 
is only true for the HSB galaxies. In the stellar and $K-$band cases, one does not 
see any clear trend. In the $B$ band, the residuals $\Delta\vm(L_B)$ correlate with 
the residuals $\Delta(B-K)(L_B)$, although mainly due to the HSB galaxies. 
The backward step--wise analysis shows that the $(B-K)$ color is indeed the third 
parameter in the $B-$band TFR (see also Kannappan et al. 2002; Pizagno et al. 2007; 
Courteau et al. 2007). There is also a weak correlation of $\Delta\vm(L_B)$ with 
$\Delta\fs(L_B)$.    

\section{Interpreting the scaling relations}

The results reported in \S 3 reveal non--negligible changes in the slopes and
scatters of the scaling correlations of disk galaxies when changing from $B$ to $K$ bands
and to stellar and baryonic quantities.  While the most robust aspects of the scaling
relations seem to be a consequence of the cosmological initial conditions of
disk galaxy formation (e.g.,  FA-R; Shen et al. 2002; Dutton et al. 2007), 
the changes found here are expected to be related mainly with the SF and gas infall 
histories of galaxies. In the following, we focus our discussion on the findings 
presented in \S 3 and on whether or not these changes are expected in \LCDM--based 
models of disk galaxy evolution.

\subsection{The baryonic relations}

The scenario of disk galaxy formation sketched in Appendix B makes certain predictions 
regarding the baryonic scaling relations of disk galaxies (Appendix C). The main simple 
physical ingredients of disk galaxy formation and evolution are the halo mass \mh, the 
mass assembly history (MAH), the spin parameter $\lambda$, and the disk galaxy mass 
fraction $\fb\equiv \md/\mh$. Below, we highlight some of the main aspects of these 
predictions with the goal of interpreting the observational inferences presented in \S 3.
At some points, we will discuss in more detail the model results obtained in FA-R. 
In that work, the cosmology used corresponded to a flat \LCDM~ universe with 
$\Omega_{M,0}=0.35$, $\Omega_{\Lambda}=0.65$, $h=0.65$, $\sigma_8=1$; \fb~ was 
assumed to be constant. 

{\bf  The \vm-\md~ relation.}
As described in Appendix C1, if the initial condition parameters $\lambda$ and \fb~ 
do not depend significantly on mass, then one expects that 
the baryonic TFR should have a slope similar to the cosmological one. {\it This 
seems to be the case according to our results} ($a_{\rm bar}\approx 0.31$ from 
observation vs $a_{\rm halo}=0.30-0.32$ from simulations). 
The models of FA-R and  Dutton et al. (2007) show that for a reasonable range
of \fb~ values (between $\sim 0.03-0.08$), changes of \fb~ with mass would not have a 
significant effect on the baryonic TFR slope (taking into account the disk gravitational 
drag on the halo); for a given \mh, as \fb~ decreases, \md~ decreases, but \vm~ also 
decreases (actually is the G function defined in the Appendix C1 that decreases), and 
the galaxy model shifts along the relation. In the same way, a 
moderate correlation (anti--correlation) of $\lambda$ with mass would only  slightly decrease 
(increase) the slope. Conversely, steep dependences of \fb~ and/or $\lambda$ upon 
mass would imply a significant change in the slope of the TFR.

Regarding the intrinsic scatter, we have measured from the observational sample a value of 
\sigintrV=0.051 for the baryonic TFR. This value, in spite
of the broadness of the sample, {\it is in marginal agreement with the theoretical
expectations discussed in Appendix C1.} 
For example, for the FA-R models with \fb=0.05, $\sigintrV\approx 0.053$; by
relaxing some model assumptions, this value is expected to be larger.
Thus, from the comparison of observations with theory, it 
seems that there is no room for other significant sources of scatter, either physical
or systematical, as for example disk ellipticity and non--circular motions (see \S 5
for further discussion).

Further pieces of information related to the scatter in the baryonic TFR are given by
the correlations between the residuals of this relation, $\Delta\vm(\md)$, 
and those of the \hb-\md, $(B-K)$-\md~ and \fs-\md~ correlations
(Fig. \ref{residuos}). The fact that $\Delta\vm(\md)$ anti--correlates 
with $\Delta\hb(\md)$ shows that {\it the radius (or surface density) plays a role in 
the scatter of the baryonic TFR}. As mentioned in \S 3.5, the radius is found to be indeed 
a statistically significant parameter in the \vm-\md-\hb~ multi--variable correlation. 
Because $\Delta\hb(\md)$ correlates strongly with \sigd~ (\S 3.5), the anti--correlation
seen in Fig. \ref{residuos} implies that, for a given \md, higher surface density disks 
produce larger values of \vm. For a given $\lambda$ distribution, the larger \fb, the higher
the disk surface density, and hence the more pronounced is the effect upon \vm~ because
the disk contribution to the total rotation curve is larger. This implies a steeper and 
stronger anti--correlation between $\Delta\vm(\md)$ and $\Delta\hb(\md)$. In the extreme
case of complete dominion of an exponential disks (no dark matter halo), this 
anti--correlation should have a slope of $-0.5$ (Courteau \& Rix 1999). 

Our observational results point to a slope 
shallower than $-0.5$, $\approx -0.15$, {\it showing that the dark halo contribution
at a radius where the rotational velocity is maximal, is significant}, as predicted by
models based on \LCDM~ halos. According to Fig. \ref{residuos}, the 
residuals are more tightly anti--correlated for the HSB sub-sample than for 
the LSB sample. This implies that the dark halo 
component becomes more dominant for lower values of the disk SB,
to the point that the contribution of the disk to the total rotation curve at radii close 
to its maximum, is negligible for the lowest SB galaxies. This dependence of the baryon 
and dark matter contents with SB has been deeply explored in ZAHF, who showed that 
for the same galaxy sample used here, only the highest SB disk galaxies are maximum disk, 
while the lower the SB, the more dark--halo--dominated (sub--maximum disk) 
the galaxies become (see also e.g., Casertano \& van Gorkom 1991; de Blok \& Bosma 2002; 
Catinella et al. 2007).  It is important to mention that the slope of the residual 
correlation is sensitive to \fb. For the modeling of disks formed inside \LCDM~ halos 
discussed above, the slope will be significantly steeper than $-0.15$ if the average value 
of \fb~ is larger than $0.05$ (e.g., Gnedin et al. 2007; Dutton et al. 2007; Pizagno et al. 
2007). For the cosmology used by FA-R and \fb=0.05, they found that the slope is 
$\approx -0.20$ (see their Fig. 8). An interesting question is {\it what slope would 
be predicted by alternative theories, like the Modified Newtonian Dynamics, and 
would agree with the value of $\approx -0.15$ found here using a broad sample of disk galaxies. }

We stress that observational studies aimed to constrain the baryon and dark matter 
contents in disk galaxies should use the baryonic relations instead of the luminous or 
stellar ones. For example Courteau \& Rix (1999) obtained the residuals of the observed 
(infrared) TFR and $R$-$L$ relation for a sample of mostly HSB galaxies, and 
interpreted the lack of correlation between these residuals as evidence of dark 
matter dominance in all galaxies. As FA-R showed, the dependency among the residuals 
changes from the baryonic to the stellar relations (see also Dutton et al. 2007;
Courteau et al. 2007). Our observational results confirm this fine effect. 

{\bf The \hb-\md~ relation.}
For the galaxy sample analyzed here, the \hb-\md~ correlation is highly scattered
(\sigintrR = 0.2 for the orthogonal fit), with the scatter indeed anti--correlating 
strongly with \sigd~ ($r=-0.91$), and less with the $(B-K)$ color,
as the \LCDM--based models predict (Appendix C2). The slope of the correlation,
for the orthogonal fit, is $0.32\pm 0.04$ (Table 2) and is steeper for the bisector
and inverse fits, i.e. is close to the slope of the CDM halo \rh-\mh~ relation 
(Appendix B). {\it Thus, according to the arguments given in Appendix C2, there is 
almost no room for a significant dependence of $\lambda$ or \fb~ on \md, in the range of
 masses of the galaxies studied here, although if both parameters anti--correlate 
(or correlate) at the same time with \md~ then the slope changes may be compensated.}

{\bf The \vm -\hb~ relation.} 
The slope of the (noisy) \vm-\hb~ correlation found here is $0.51\pm 0.11$ (orthogonal fit), 
i.e. shallower than the cosmological one (see Appendix B). This, according to Appendix C3
{\it favors, if any, an anti--correlation of \fb~ with mass}. The intrinsic scatter, 
\sigintrV=0.16, is much larger than the scatter around the baryonic TFR.  As predicted by 
the models, the residuals of the \vm-\hb~ correlation correlate significantly with both 
log\sigd~ and the $(B-K)$ color ($r=0.95$ and 0.67, respectively); the slopes of these 
correlations (orthogonal fit) are 0.34 and 0.28, respectively.

\subsection{The stellar relations}

The main changes when moving from the baryonic to the stellar scaling correlations 
are the decrease of the slope and intrinsic scatter in the \vm-$M$
correlations. The scatters of the $R$-$M$ and $R$-\vm~ correlations
also decrease but the significance of these decrements are very marginal.
More interestingly, the (weak) anti--correlation between the residuals of the baryonic
\vm-\md~ and \hb-\md~ correlations tends to disappear in the stellar case.
This is a fact related to other of our results, namely that {\it the radius (or disk 
surface density) is a third parameter in the baryonic TFR but no longer in the stellar one}.  
All these differences are related to the gas (or stellar) mass fraction of galaxies, and they show that the mass infall and SF histories vary systematically among
galaxies. Are these variations consistent with galaxy models based on the \LCDM~ framework?

\begin{figure*}
\vspace{8.5cm}
\hspace{15.3cm}
\includegraphics{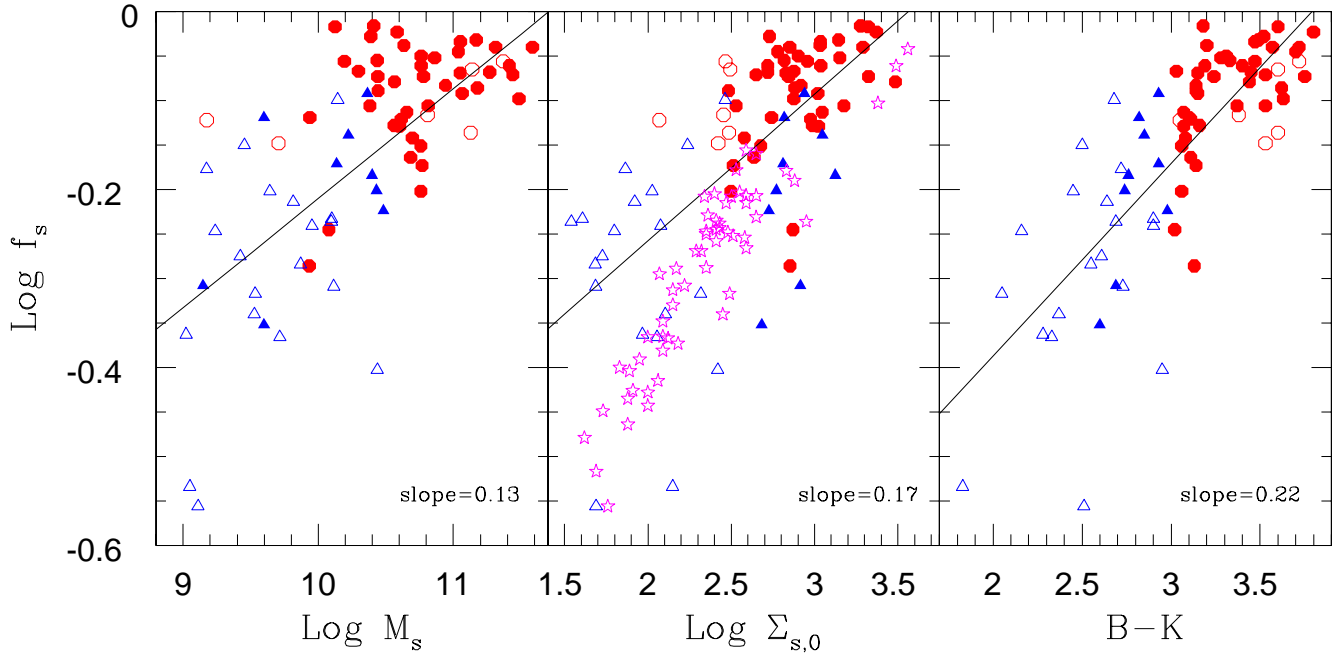}
\figcaption{Stellar fraction \fs~ vs \ms, \sigs, and $(B-K)$. Solid and empty symbols are for 
HSB and LSB galaxies, respectively. Red galaxies are represented with 
circles, while blue galaxies with triangles. The solid lines are the Log--Log orthogonal 
linear regressions.  Stars are from models by FA-R.
\label{stfrac}
}
\end{figure*}

Let us first discuss the change in the slope of the \vm-$M$ relation.  
If the stellar mass fraction, $\fs=1-\fg$, depends on \ms~ as $\fs\propto \ms^{\beta}$, 
then $\md=\ms/\fs\propto \ms^{1-\beta}$. Therefore, from $\vm\propto \md^{\alpha}$
one passes to $\vm\propto \ms^{\alpha (1-\beta)}$. In the lesft panel of Fig. \ref{stfrac}, 
we show \fs~ vs \ms~ for our galaxy sample. Although with a high scatter, \fs~  
correlates with \ms~ roughly as 
\begin{equation}
\fs = 0.65 (\ms/10^{10}\msun)^{0.13}, 
\end{equation}
implying then that the slope from the baryonic to stellar TFRs should decrease by a 
factor $\sim 0.87$. The corresponding slopes calculated with the orthogonal fits are 0.31 
and 0.27, respectively. What causes the dependence of \fs~ (or \fg) on \ms? It could be
that rather than a fundamental dependence, it is a consequence of other correlations. 
For example, Fig. \ref{stfrac} shows that \fs~ correlates stronger with \sigs~ 
(and also with galaxy color) than with \ms. 

From the point of view of the models, the stellar fraction \fs~ of normal disk galaxies 
depends mainly on two factors: the efficiency of gas transformation into stars, given 
basically by the disk surface density (determined by $\lambda$, \fb, and \mh), and the gas 
infall history, connected to the halo MAH. Based on these processes and by means of 
semi--numerical models, FA-R and Avila-Reese \& Firmani (2000) have shown that (i) lower 
surface density disks transform gas into stars with less efficiency than disks of high 
surface density, (ii) more massive disks tend to be of higher surface density (see also 
Dalcanton et al. 1997), and (iii) the present--day \fs~ correlates significantly 
with \sigs~ (or \sigB), and because
of item (ii), with \ms. Note that the dependence of \fs~ on \sigs~ is a non trivial 
prediction because several physical/evolutionary processes combine to give rise to the 
stellar (or gas) mass fraction and surface density (brightness) of evolved disk galaxies.
In Fig. \ref{stfrac}, together with the observational inferences, the model predictions 
presented in FA-R are plotted in the \fs-\sigs~ diagram (magenta stars). 
Models and observations occupy the same region, though the models have on average
slightly lower \fs~ and \sigs~ values; the introduction of interaction--driven
SF would work in the direction of increasing both quantities.

As a consequence of the mentioned predictions, FA-R showed that the scatter around the 
stellar TFR due to $\lambda$ (\sigd) should diminish compared to the one around the 
baryonic TFR; from the baryonic to the stellar TFR, the average  
scatter decreases from $\sigintrV\approx 0.053$ to $\sigintrV\approx 0.040$. 
Galaxies that in the baryonic
TFR were shifted to the high--\vm~ side, because of their higher surface densities, 
transform gas into stars more efficiently and shift also to larger \ms~ values in the
stellar TFR diagram compared to galaxies of lower surface densities. As a result,
the dispersion in the stellar TFR due to disk surface density (mainly associated
to $\lambda$ and \fb) decreases, and {\it the TFR becomes almost the same for HSB and
LSB galaxies: the anti--correlation between the residuals of the TFR 
and those of the \hs-\ms~ relation disappears.}
The dispersion in the halo MAH remains as the main source of scatter in the stellar TFR.
However, regarding the original MAH dispersion, it is also predicted that there is a 
kind of compensation in the shift of galaxies in the \vm-$M$ diagrams: 
galaxies that were shifted to the high--\vm~ side because they were formed in 
halos assembled earlier (more concentrated), are also shifted 
to the larger--\ms~ side because these galaxies had more time to consume their gas 
(they end up with higher \fs). 

The predictions of the \LCDM--based disk galaxy models provide an adequate context
for interpreting the observational results plotted in the two upper rows of Fig. \ref{residuos}.
The loss of dependence of the \vm-$M$ correlation residuals upon the $R$-$M$ correlation 
residuals when changing from baryonic to stellar quantities can be understood on the ground of 
the (self--regulated) SF efficiency dependence on disk surface density 
described above (see also Dutton et al. 2007).  
This also explains the reduction of the scatter around the TFR from the baryonic case to the
stellar case (see Table 1), though the model results are more pronounced than observations 
both in the changes of the scatter values and in the slopes of the residual correlations.

In the same way, the distribution of galaxies in the diagrams of the residuals for the 
\vm-$M$, $(B-K)$-$M$, and \fs-$M$ correlations (Fig. \ref{residuos}) can be interpreted 
as follows. Upon the understanding that disk color and \fs~ are related mainly to the 
halo MAH (concentration), the correlations between the baryonic residuals $\Delta\vm(\md)$ 
and $\Delta(B-K)(\md)$, and between $\Delta\vm(\md)$ and $\Delta\fs(\md)$, are expected 
(see above). This is because for a given mass, 
halos assembled earlier have larger \vm, and the galaxies formed inside them transformed gas 
into stars earlier (therefore are redder) and more efficiently (therefore have higher values 
of \fs). In the corresponding panels of Fig. \ref{residuos}, 
one indeed sees these predicted tendencies, mainly for the HSB galaxies. For the LSB 
galaxies, the tendencies again (see \S\S 3.4) almost disappear. It might be 
because the (low) SF
rate in LSB galaxies no longer depend on the gas infall rate; this dependence
is what allowed us to associate the observables, color and \fs, to the halo 
MAH\footnote{Below a certain gas surface density, the SF 
rate could be so inefficient that its long scale--time becomes independent of the gas 
infall rate (given by the MAH). Alternatively, the SF rate in LSB galaxies could be 
disconnected from the gas infall rate and the gas surface density as a consequence of their 
low metal content, which makes the gas cooling very inefficient (Gerritsen \& de Blok 1999).}.
In the residual plots corresponding to stellar quantities, the tendencies become negligible 
because from \md~ to \ms, the shifts in \vm~ in the TFR, due to the halo MAH 
(concentration), tend to be compensated by shifts in \ms~ (see the previous paragraph).
Therefore, {\it in the \vm-\ms~ (stellar) diagram the mentioned SF effects tend to
reduce the scatter and its dependence on surface density, color, and \fs~ compared to 
the \vm-\md~ (baryonic) diagram.}

\subsection{The luminous relations}

The slopes and scatters of the $K-$band scaling correlations are similar,
within the uncertainties, to those of the stellar scaling correlations. The only quantity 
that actually changes in the diagrams is $L_K$ for \ms, through the $K-$band 
mass--to--luminosity ratio, $\Upsilon_{K}$. This ratio increases slightly with the 
disk color for HSB galaxies and slightly decreases with it for blue LSB galaxies 
(\S 2.2). Hence, a significant trend of $\Upsilon_{K}$ with mass 
or \vm~ is not expected. For the scale length, we assumed that \hs=\hK. 

Interesting differences are seen between the $K-$ and $B-$band scaling correlations.
In the \vm-$L$ diagrams, the slope and intrinsic scatter increase significantly 
from the $K$ to the $B$ band. Calculating the ratio of the $K-$ and $B-$band TFRs, 
we obtain: $(B-K)\propto \vm^\delta$, where $\delta$ is $-2.5$ times the difference 
of the $B-$ and $K-$band slopes. This correlation is a consequence of the 
color--$L$ correlation shown already in Fig. \ref{colorSB}. That the less luminous galaxies
systematically tend to be slightly bluer, implies already a steepening of the 
slope when changing from the $K-$ to the $B-$band TFR. 
 
Furthermore, the residuals of the $B-$band TFR are correlated with those 
of the $(B-K)$-$L_B$ correlation, and less with the residuals of the \hB-$L_B$ 
and \fs-$L_B$ correlations: redder galaxies, with higher SB and \fs~ 
tend to be shifted to the TFR large--\vm~ side; the opposite happens for bluer galaxies, 
with lower SB and \fs~ (lowest row panels in Fig. \ref{residuos}).  
In fact, the step--wise backward analysis mentioned in \S\S 3.5 shows that the 
$(B-K)$ color is a statistically significant parameter in the \vm-$L_B$-$(B-K)$ multi--variable 
correlation, i.e. {\it color is the third parameter in the $B-$ band TFR,} 
a result that other authors have also found in the optical bands 
(e.g., Verheijen \& Sancisi 2001; Courteau et al. 2007; Kannappan et al. 2002, and 
more references therein). It is obvious
that a galaxy in the \vm-$L_K$ diagram shifts to the large (low) 
$B-$band luminosity side if it is bluer (redder) than the corresponding average. 

Why do galaxies shifted to the high-- (low--)\vm~ side in the
$B$ band TFR tend to be redder, with higher SBs and larger values of \fs~ 
(bluer, with lower SBs and smaller values of \fs)?  According to the galaxy 
models discussed above, disks formed in halos
with low $\lambda$ and/or early mass assembling (high concentration), tend to
consume gas into stars early in such a way that, for the same \ms~ or $L_K$,
they result with older stellar populations (lower $L_B$ and redder) and a higher 
\fs~ than those formed in halos with high $\lambda$ and/or late mass assembling 
(low concentration). Thus, {\it while the dispersions 
introduced by $\lambda$, \fb, and the MAH are reduced by compensating effects in the stellar 
and $K-$band TFRs (\S\S 4.2), the opposite happens in the $B-$band TFR}. We see indeed  that 
the TFR in blue bands is systematically more scattered than in the NIR bands (Table 1).

Regarding the \hB-$L_B$ correlation, its slope is significantly steeper than that
of the \hK-$L_K$ correlation. Again this is because, as the galaxy is more luminous,
it tends to be redder, i.e., the $L_B$--to--$L_K$ ratio decreases with $L$.
However, the situation is more complex than in the TFR case, because
in the $R$-$L$ diagrams (i) the scatter is large and galaxies are strongly segregated by 
SB and color, and (ii) the $B-$ and $K-$band radii are not the same as is the case of \vm~ in 
the TFR diagrams; the \hB--to--\hK~ ratio tends to be larger for redder, more luminous galaxies.
Thus, on one hand, the redder, higher SB galaxies shift to the low--$L$ and high--$R$ sides 
in the $R$-$L$ diagram when changing from band $K$ to band $B$; the shift tends to be larger 
for the more luminous galaxies (see Fig. \ref{RL}). On the other hand, the bluer, lower 
SB galaxies shift slightly, mainly to the low--$L$ side, while the shift is larger for those 
of larger luminosities. As a consequence, the slope of the correlation increases in the $B$ 
band and the scatter along the luminosity axis decreases. 

In the \vm-$R$ diagrams, the correlations do not significantly change when changing from 
the $K$ to the $B$ bands. This confirms that the most relevant shifts in the
$R$-$L$ diagrams are due to $L$ rather than $R$. The models predict that the scatter around
the \vm-$R$ relation should thicken slightly in the band $B$ with respect to the band $K$: 
galaxies of lower
SBs and/or bluer (larger $\lambda$, late assembling halos, and lower \fb), which are 
shifted to the low-\vm~ side in the $K$ band, tend to shift also to larger $B-$band
radii with respect to their \hK~ radii; for galaxies that are of higher SB and redder, the
radii tend to be the same in both bands.  According to Table 3, the direct
intrinsic scatter around the \vm-\hK~ and \vm-\hB~ correlations is almost the same,
while the inverse scatter (along the $R$ axis), is larger in the $B$ band.

\subsection{Quantitative comparisons}

As shown, the predictions obtained with the \LCDM--based models by FA-R 
offer an excellent qualitative description of several observational details of isolated 
disk galaxies. A systematic comparison of observations with results from models calculated 
with the most recent cosmological parameters will be presented elsewhere. 
With the aim to show that the models are able to provide even quantitative predictions, 
we plot those results from FA-R that can be compared with the observational inferences 
presented here. In FA-R, the authors plotted averages and scatters for three mass bins 
in the stellar TFR diagram (their Fig. 7), that we reproduce here in the corresponding 
panel of Fig. \ref{TF} (magenta stars with vertical error bars). We diminished their \ms~ 
points by the factor (0.70/0.65)$^{-2}$ in order to take into account the slightly different
value of $H_0$ used in their models. The agreement is good, even more if we consider that 
models calculated with a $\sigma_8<1$, would have smaller values
of \vm. The agreement is also quite good in the baryonic TFR diagram (magenta stars in the lower 
right panel of Fig. 1; the \fg~ values given in FA-R were used to calculate \md~ from \ms).
It should be stressed that models and observations are now being fairly compared; in both cases
we have nearly isolated normal disk galaxies in a wide range of SBs and morphological types.

 In FA-R, model results for the \hs-\vm~ relation were also presented, 
separated in HSB and LSB galaxies, but models with very high SB were not taken
into account. We estimate the ``weight'' of these models in the fit for the HSB galaxies, 
and reproduce both the LSB and HSB+very HSB fits in the corresponding panel of Fig. \ref{VR} 
(thick magenta dashed lines), correcting the radii by a factor of (0.70/0.65)$^{-1}$.  
The agreement is rather good within the large scatter.  
Finally, model results by FA-R in the \fs-\sigs~ diagram
(Fig. \ref{stfrac}, magenta stars), and (ii) in the diagram of residuals of the baryonic 
\vm-\md~ and \hb-\md~ correlations (upper left panel of Fig. \ref{residuos}, magenta dashed 
line) were already presented above, evidencing a reasonable agreement with observations; 
for the stellar case, the residuals do not correlate, as seen also for the observations.

\section{Possible difficulties for the disk \LCDM--based models}

The hierarchical \LCDM--based models of disk galaxy evolution discussed above have been 
useful for interpreting the results reported in \S 3 about the slopes and scatters of the 
observed scaling 
relations in the different cases, as well as the behaviors of the residuals from these and 
other correlations. Even the zero--points of the relations predicted by FA-R seem to agree 
quantitatively with observations. Note that the models by FA-R, differentiate from others 
in the literature in that they {\it include a detailed description of disk SF and feedback 
and follow self--consistently the whole process of halo--disk evolution}. Although a detailed 
comparison with observations, using models calculated with more recent cosmological parameters, 
was left for a forthcoming paper, we may anticipate here some potential shortcomings of the 
model predictions as well as solutions. 

(1) {\it The models are not able to explain the observed correlation of galaxy color 
with mass or luminosity}. In the 
hierarchical scenario, isolated more massive halos tend to assemble a given fraction of 
their present--day mass later than the less massive ones. Then disks formed inside the more
massive halos would tend to be younger (bluer) than those formed in the less massive halos. 
However, the models also show that the more massive the halo is, the higher the surface 
density of the disk formed inside it (Dalcanton et al. 1997; FA-R); the higher the disk 
surface density, the more efficient the SF process and therefore, the redder the galaxy. 
As a result of these two evolutionary compensating effects, the models predict
almost no dependence of color on $M$ or $L$ (e.g., Avila-Reese \& Firmani 
2000; Avila-Reese et al. 2005, see Table 5 therein). A possible way to produce a steeper
color--$M$ correlation in the models is by obtaining a steeper disk surface density--$M$ 
dependence; in fact, the model prediction for this dependence is also too shallow as 
compared with observations (see below). It should be said that the models do not 
include satellite galaxies
and interacting--induced SF. For massive systems it is probable that the
late gas infall is reduced by the large cooling time and because a fraction
of the gas could have been trapped before into substructures that tend to be incorporated into the 
central galaxy through (minor) mergers; all these processes make the most luminous
galaxies redder and with larger \fs~ values than obtained in the FA-R models.

An indirect dependence of galaxy color on $L$ may also appear through environment. 
However, the disk galaxy sample studied in this paper includes field galaxies and galaxies 
from the UMa low--density, low mass cluster of galaxies. Therefore, it is less probable that 
environmental effects should produce the color--$L$ correlation. This correlation, which in 
some sense resembles the so--called {\it anti--hierarchical or downsizing galaxy formation} 
reported for early--type luminous galaxies (e.g., Cimatti, Daddi \& Renzini 2006), 
could be produced by a mass (luminosity)--dependent dust extinction larger than that used  
here to correct the observations. Another alternative is related to the AGN feedback mechanism 
invoked recently to solve the sharp cut--off at the high--end of the (elliptical) galaxy
luminosity function, as well as the observed downsizing of red early--type galaxies (e.g.,
Bower et al. 2006). Such a mechanism should be, however, strongly dependent on the galaxy mass
in order to produce the observed continuous and monotonic change of color with mass (luminosity)
of disk galaxies. 

(2) {\it The predicted dependence of surface density on mass (or SB on luminosity) is too 
shallow as compared to that inferred from observations}, although the latter is very noisy. 
Such a dependence can be estimated by combining the two simple model relations: 
\begin{eqnarray}
\hb\propto \lambda\ g(c) \rh\ \propto \lambda\ g(c)\ (\md/\fb)^{1/3}, \\ \nonumber
\hb\propto (\md/\sigd)^{1/2}. 
\end{eqnarray}
By rejecting the small dependence of $g(c)$ on mass (see Appendix C2), we then obtain that
\begin{equation} 
\sigd\propto \md^{1/3}\fb^{2/3}/\lambda^2. 
\end{equation}
The semi--numerical evolutionary models show that the dependence of \sigd~ on \md~ is 
even shallower than 1/3. On the other hand, the slope 
(orthogonal regression) that we have found for the (noisy) correlation between \sigd~ and 
\md~ in our galaxy sample is $\sim 0.7$, and it does not change significantly for 
the stellar and luminosity cases. 
Assuming that there is no a dependence of \fb~ and $\lambda$ on mass, model 
predictions and observations would not agree: the surface density (brightness) of observed 
galaxies increases on average more rapidly with $M$ ($L$) than the models predict. 
Again, the inclusion of mergers and interaction--induced SF would work in the direction
of increasing \sigs~ in the more massive galaxies.

On the other hand, recent inferences of the halo $\lambda$ 
distribution for a large SDSS sub--sample of galaxies show this parameter to be systematically 
smaller and less scattered for more luminous galaxies (Cervantes-Sodi et al. 2007; Berta et al. 2008;
see also Hernandez et al. 2007). By taking into account an anti--correlation of $\lambda$ 
with mass in the models, both the color (1) and surface density (2) problems could be 
ameliorated without affecting significantly the scaling correlations. The question is 
why more massive galaxies would end with smaller
values of $\lambda$ --and consequently, higher surface densities-- than the less massive
ones.  Cervantes-Sodi et al. (2007) suggested that the trend of $\lambda$ with mass can
be related to the halo angular momentum acquisition. 
A complementary explanation may lie in the baryon processes. For example, the outer gas 
in the most massive
galaxy halos may not have enough time to cool and fall to the disk (for numerical
results see e.g., van den Bosch et al. 2002; Keres et al. 2005);  the material 
located in the outer regions is the richest in the specific angular momentum 
(Bullock et al. 2001b). On the other 
hand, in less massive halos the outer gas may flow through filaments to the
disk, even before being shock--heated by the halo collapse (Keres et al. 2005). As 
a result, the specific angular moment content will be 
higher in low mass galaxies and lower in the massive ones, which implies lower and
higher disk surface densities, respectively. Note that this scenario also implies
that more massive galaxies would tend to have lower values of \fb.  A mild 
anti--correlation of \fb~ with mass in the models is allowed according to the 
observational results presented here (\S\S 4.1): 
the \vm-\hb~ relation would agree with the anti--correlation, while the \hb-\md~ 
relation would do so in case $\lambda$ also anti--correlates with mass, and 
regarding the \vm-\md~ relation, there is room for a mild variation of 
\fb~ with mass. 
Detailed model calculations are necessary for obtaining more quantitative 
conclusions.

(3) {\it Are the scatters around the model and observed TFRs in agreement?}
For the baryonic case, the agreement is marginal, given the scatter in the models 
is slightly larger than that in the observations (\S\S 4.1). The difference could 
be even larger if (i) some of the simplifying assumptions introduced in the 
models are relaxed, which would
increase the predicted scatter, and/or (ii) systematical errors such as disk ellipticity 
are present, which would give rise to a reduction in our estimate of the observational 
intrinsic (physical) scatter%
\footnote{Some researchers have found evidence that disk ellipticity could account for roughly 
50\% of the observed scatter in the luminous TFR (Andersen et al. 2001; see also Franx \& de 
Zeeuw 1992;  Rix \& Zaritsky 1995) or it could even explain all the observed scatter (Ryden 2004). }. 
Therefore, the current models of disk galaxy formation based on the \LCDM~ cosmology have a 
potential problem in predicting the intrinsic scatter around the baryonic TFR. This
scatter is smaller in the models for lower values of \fb~ or for a distribution of 
$\lambda$ narrower than that of the \LCDM~ halos. It could also be possible that the 
potential systematic errors, such as disk ellipticity, confabulate with the disk 
surface density or baryonic--to--total mass ratio (with $\lambda$ and \fb~ from the  
model's point of view) in such a way that the observed TFR scatter is 
systematically diminished.

For the stellar case, the scatter in the models becomes slightly smaller than that 
in the observations (\S\S 4.2). Although in both cases (models and observations), 
the scatter decreases from the baryonic TFR to the stellar TFR, for models, the 
decrease is more pronounced than it is for observations. However, there are known 
effects not considered in the FA-R models, like the interaction--induced 
SF, that will work certainly in the direction of increasing the scatter 
around the stellar TFR  (e.g., Barton et al. 2007), but not around the baryonic one.

(4) {\it Are current models able to match both the TFR zero--point and the galaxy luminosity function?}
First, we note that the disk galaxy evolutionary approach, like the one of FA-R, is 
different by construction to the the semi--analytic models (SAM) or the halo occupation 
models (HOM): the former is 
focused on the modeling of the galaxy internal physics and evolution, while the latter 
are constructed for reproducing galaxy population global properties as the luminosity function. 
The parameters that the SAM and HOM need for reproducing in detail the observed luminosity 
function apparently produce a luminous TFR whose zero--point is shifted to the higher--velocity 
side with respect to observations (Baugh 2006, and references therein). Some astrophysical 
solutions were proposed to alleviate this apparent problem (e.g., Dutton et al. 2007; Gnedin 
et al. 2007). 

However, before introducing modifications to the common disk galaxy formation 
scenario, we may mention at least four aspects that should be taken into account in order to 
compare fairly the predictions of SAM/HOM with observations in regards to the TFRs. 
(1) Since the TFRs are related to internal disk properties, a more detailed modeling of 
these properties than that carried out in the SAM is necessary; for example, in the SAM, 
disks are assumed to have a purerely exponential mass 
distribution, the SF--feedback processes are artificially linked to the halo circular velocity, 
etc. (2) \vm~ is typically used to define the TFRs, while in the SAM/HOM only approximations to 
this velocity are calculated. (3) The TFRs
refer only to the population of normal disk galaxies, while the luminosity function 
that the SAM/HOM fit to observations refers to the overall galaxy population. (4) The halos
where disk galaxies form could be a sub--group of the overall halo population, for example
those that did not suffer a major merger since $z\sim 1$; therefore, the corresponding 
``disk galaxy'' halo mass function, and not the overall one, should be used in the HOM 
as the starting point in the procedure of shaping the luminosity function.

From the point of view of the SAM and HOM, the passage from the halo mass function 
to the galaxy luminosity function implies a specific \mh--to--$L$ ratio as a function 
of mass. Recent studies using galaxy--galaxy weak lensing, allowed one to directly 
determine the 'central' galaxy \mh--to--$L$ ratios (e.g., Hoekstra et al. 2005; 
Mandelbaum et al. 2006). For the FA-R models, which would agree very well with the 
baryonic and stellar TFRs if corrected for $\sigma_8<1$, we have the $B-$band 
luminosities available. The results from Mandelbaum et al. (2006) for {\it late--type} galaxies (note that they differ from 
those of the early--type galaxies) are given in the $r$ band. By using $<(B-R)>=1.2$ 
for late--type galaxies (de Jong 1996b) and $(r-R)=0.2$, we estimate from the
FA-R models $<\mh/L_r>$/(\msun/$L_{r\odot}$) = $62.2\pm9.1, 60.7\pm14.3,$ and $77.6\pm26.5$,
for the luminosities $<L_r>/L_{r\odot} = (5.6\pm1.1)\times 10^{10}, (5.7\pm1.2)\times 10^{9},$ 
and $(4.85\pm 1.2)\times 10^{8}$, respectively. To compare the FA-R $\mh/L_r$ ratios with
those of Mandelbaum et al., the values just reported should be increased 
by $\approx 15\%$ due to the different definitions of \mh. The good agreement between
the models and observations within their scatters is encouraging. 
The model predictions also agree roughly with the $<\mh/L_r>$ and $<\mh/\md>$ 
ratios calculated from the observed luminosity function and the stellar
mass--to--$L$ ratios inferred from kinematical mass modelling (Shankar et al. 2006).
We can also compare tracers of the inner luminous (or baryonic)--to--dark matter 
ratios of models and observed galaxies. For example, in ZAHF the disk--to--total 
velocity ratio at the maximum of the rotation curve was estimated for the same 
galaxies studied here. The models and observations were found to agree roughly, 
both in the zero point as well as in the dependence of this ratio on the disk 
surface density or SB (see Fig. 3 in ZAHF).

Summarizing, the FA-R models based on the \LCDM~ scenario seem to be able to 
predict both the TFR, including its zero point, and the inferred, directly 
from observations, \mh--to--luminosity ratios of central late--type galaxies, as well 
as the inner luminous and baryonic--to--dark matter contents. These ratios, specially the 
former, are closely related to the ``transfer'' function needed for passing from the
halo mass function to the galaxy luminosity function. We speculate that if the SAM and HOM
take into account the shortcomings mentioned above, then the potential problem of
fitting both the TFR zero point and the luminosity function will be overcome. On the other
hand, some of the initial condition parameters used in models such as those of FA-R, if any, 
could be modified accordingly to reproduce the disk--galaxy luminosity function, while 
the TFRs (and the other scaling relations) would be yet in agreement with the observations.
For example, the  anti--correlation of \fb~ with mass for luminous galaxies (motivated by the 
large cooling time in massive halos) that the SAM/HOM evoke, could be allowed by the models
as discussed in items 1 and 2 above.

\section{Summary and Conclusions}

A compiled sample of 76 normal (non--interacting) local disk galaxies of all 
morphological types and SBs has been used to construct and compare, among them, the luminous 
(bands $B$ and $K$), stellar and baryonic scaling relations. The required observational 
information for building up this sample implied detailed surface photometry in bands $B$ 
and $K$, a dynamical determination of the rotation curve amplitude, and HI gas integrated 
flux data. We have corrected and processed the observational data homogeneously, and 
obtained several composite disk quantities and their uncertainties: global $(B-K)$ colors,
stellar masses and surface densities, gas (or stellar) fractions, and baryonic masses, surface
densities and scale lengths. The objectives of this work were (i) to explore the changes of 
the disk--galaxy scaling relations and their residuals as moving from optical to NIR bands, 
and to stellar and baryonic quantities, (ii) to look for interpretations of the results 
in the context of \LCDM--based evolutionary disk--galaxy models, and (iii) to show the necessity
of larger samples, similar to the one compiled here, for constraining models of galaxy formation and
evolution.  A summary of the results is as follows.

-- The slope of the baryonic TFR is shallower than the slopes of the stellar and $K-$band TFRs
($\approx 0.31$ vs $\approx 0.27$ and 0.26; orthogonal regression) and, in agreement with 
previous works, the slope of the $B-$band TFR is steeper than that of the $K-$band TFR 
($\approx 0.32$ vs $\approx 0.27$). The estimated average intrinsic scatters in log\vm~ 
around the baryonic, stellar, $K-$ and $B-$band TFRs are 0.051, 0.045, 0.049, and 0.063 
dex, respectively. Thus, the baryonic TFR is more scattered than the stellar TFR. A statistical 
analysis shows that the radius (or surface density) is a significant third parameter in the 
former relation, while the latter, as well as the $K-$band TFR, do not admit a third parameter.  
In the case of the $B-$band TFR, which is the most scattered, the $(B-K)$ color is a 
significant third parameter.

-- The $R$-$M(-L)$ correlations are scattered and segregated strongly by disk SB and weakly by
color. The slopes of the baryon, stellar, and $K-$band correlations are around 0.30 (orthogonal 
regression), while for the $B-$band relation, this increases to 0.44. The scatter around the 
latter, projected in the luminosity axis, is smaller than that in the other correlations. The 
$\vm-R$ correlations are the most scattered and are also segregated strongly by disk SB and 
less by color. The slopes have values of around 0.5.

-- The residuals of the baryonic \vm-\md~ and \hb-\md~ correlations are 
anti-correlated, mainly for the HSB galaxies. The slope is $-0.15\pm 0.04$ (orthogonal 
regression), showing that the smaller the disk radius, for a given mass (higher surface density), 
the larger the disk contribution to \vm. For the stellar and $K-$band cases, the anti--correlation 
almost disappears.  Thus, our results show that the correlations among the mentioned residuals are 
different for the baryonic, stellar, $K-$ and $B-$band cases.   
While the correlation (or the lack of it) among the residuals in the baryonic case can be used 
to explore the dynamical importance of the halo/disk in galaxies, in the stellar
and luminous cases the effects related to SF processes distort any pure dynamical
interpretation of the results.       

-- In spite of the velocity--dependent internal extinction correction that we have applied,
the $(B-K)$ color correlates significantly with the logarithms of $L_B$, $L_K$, \ms, and \md\ 
(orthogonal regression slopes of 0.61, 0.53, 0.56, and 0.60, respectively). Noisy but steep 
correlations are observed between the surface densities and brightnesses and the corresponding 
masses and luminosities. The stellar fraction, \fs, correlates strongly with $(B-K)$ 
and less with surface densities (brightnesses) and 
luminosities (masses).  In all cases, the lowest SB galaxies break the latter correlations, 
suggesting a kind of threshold in the surface density, below which the SF 
rate is almost independent of it (and of $M$ or $L$) and of the gas infall rate.

We have discussed previous models of disk--galaxy formation and evolution (including halo
contraction, SF, and disk feedback; e.g., FA-R) within hierarchically growing \LCDM~ halos, under 
the assumption of gas--detailed angular momentum conservation. We showed the potentiality of 
these models to describe the observational inferences, as well as how such inferences can 
help us to constrain several of the model parameters. 
A crucial aspect of this endeavor is the correct comparison of models with observations. The
basic models refer to isolated normal disk galaxies. Therefore, the observational sample 
should also be for nearly isolated normal disk galaxies.
We highlight the following conclusions:

$\bullet$ The slopes of the baryonic \vm-\md, \hb-\md, and \vm-\hb~ correlations can be 
interpreted as a direct imprint of the cosmological \LCDM~ halo \vmh-\mh, \rh-\mh, and \vmh-\rh~ 
relations, especially the former (baryonic TFR), which is the tightest one.
The models show that this relation is robust to systematical or 
statistical variations in the disk baryonic mass fraction, \fb, for $\fb<0.08$. 
Then, the main sources of scatter are the dispersions in the halo 
spin parameter $\lambda$ and halo MAH (or concentration). The FA-R models provide a 
lower limit to the scatter around the relation due to these two dispersions, namely 
\sigintrV$\approx 0.053$ (\fb=0.05=const.), which is already slightly larger than
our observational inference. 

$\bullet$ In combination, the slopes and scatters of the baryonic \vm-\md, \hb-\md, and 
\vm-\hb~ correlations inferred here from observations, as well as the correlations among 
their residuals, imply that the average value of \fb~ in normal disk galaxies could
not be larger than $\sim 0.05$ and that \fb~ can not depend strongly on $M$ (or $L$); 
if any, a moderate anti--correlation with $M$ is permitted. 
A trend of \fb~ nearly independent on mass, with average values of $\fb<0.05$,
has been reported recently by other authors, who extended the analysis to disk dwarf galaxies 
and inferred halo masses directly from the observational data
(Blanton et al. 2007; see also Baldry et al. 2008; Begum et al. 2008).

$\bullet$ The models are able to explain qualitatively the observed (weak) anti--correlation 
among the residuals of the {\it baryonic} \vm-\md~ and \hb-\md~ correlations and the related 
finding that the radius (or surface density) is a third 
parameter in the former correlation. The physical parameter beyond these behaviors is $\lambda$. 
Model disks with low $\lambda$ have
high surface density and contribute significantly to \vm~, in such a way, that for a given mass, 
\vm~ correlates with \hb~ or \sigd.  Instead, disks with high $\lambda$ values 
are of low surface density and have a negligible gravitational contribution to \vm; then, 
the dependence of \vm~ on \sigd~ (or on \hb~ for a fixed mass) tends to disappear. 
The models can also explain 
why the anti--correlation among the residuals becomes negligible in the stellar or 
$K-$band cases: for a given mass, galaxies not only shift to the high-\vm~ side in the TFR 
diagram as the radius is smaller (\sigd~ is higher), but also shift to the 
high-\ms~ (or $L_K$) side due to a higher efficiency in transforming gas into stars. 
As a result, the scatter around the stellar TFR becomes independent of \hs~ or \sigs, and
smaller than in the baryonic case.  

$\bullet$ The (self--regulated) SF efficiency effect just mentioned produces a dependence 
of the stellar mass fraction \fs~ on surface density (or brightness). Such a non--trivial
prediction roughly agrees with the one found for our processed galaxy catalog 
(Fig. \ref{stfrac}), and it is associated mainly with the $\lambda$ parameter: the smaller 
the $\lambda$, the higher the disk surface density and the more efficient the transformation
of gas into stars. Although less relevant, \fs~ and \sigs~ are also affected by the gas 
infall history, which is connected to the cosmological halo MAH. Since the 
MAH plays a role in the scatter of the baryonic TFR, trends among the residuals of this 
relation and those of the color-\md~ and \fs-\md~ relations are predicted. These trends were found 
for our observational catalog, mainly for the HSB galaxies (Fig. \ref{residuos}). For LSB 
galaxies, it could be that the SF--rate timescale becomes larger than the that of the gas 
infall rate, the connection between MAH and color/\fs~ then disappears.

$\bullet$ {\it Potential difficulties.-} 
Our observational results show that both (i) the color--$M$ and (ii) surface density--$M$ 
correlations are steeper than the models would predict. We speculate that these
shortcomings suggest that late minor mergers and interaction--induced SF, which are more 
common in massive galaxies, should play some role and/or that both, the galaxy 
spin parameter and 
\fb~ anti--correlate with mass; at least qualitatively, both anti--correlations, in 
combination are not expected to affect the scaling correlations presented here.
(iii) The fact that the intrinsic scatter around the predicted baryonic TFR appears to be
larger than that inferred here, implies that there is no room for other physical or 
systematical sources of scatter around the TFR such as disk ellipticity. On the other hand,
the intrinsic scatter around the stellar TFR becomes smaller in the models than in the
observations. However, external effects, such as the interaction--driven SF, work in the 
direction of adding some scatter to the stellar TFR.
(iv) The FA-R models (rescaled to $\sigma_8<1$) agree with the observed baryonic 
and stellar TFR zero points as well as with the \mh--to--$L$ ratios inferred directly 
from observations. The potential difficulty that the SAM/HOM find in matching both the 
TFR zero point and the luminosity function could be solved by taking into account the 
shortcomings we have discussed in \S 5.

The results obtained in this work pose relevant questions
for future studies. We showed that the determination of the scaling relations for disk galaxies 
in different bands and for stellar and baryonic quantities, as well as for the correlations 
among their residuals, bring valuable information on the nature and evolution of galaxies.
The completion of large galaxy samples, unbiased and broad in properties, with detailed 
photometric, dynamic, and gas--content information is crucial for this kind of study.
The preferred approaches for such an undertaking are: follow--up observations
in $H_\alpha$ and/or HI emission lines for galaxies from the largest optical/NIR band
surveys as SDSS (Pizagno et al. 2005), and synergy of existing wide--area homogeneous surveys 
conducted at optical/NIR bands and HI emission line, for example SDSS with HI surveys 
such as ALFALFA
(Giovanelli et al. 2005).  According to the philosophy followed in this paper,
substantial advance comes from the adequate comparison of observations with 
theoretical models. Thus, in order to confirm, rule out, or perhaps expand, some of the 
conclusions attained here, a deep model exploration of the formation and 
evolution of the disk galaxy population should be carried out.

\acknowledgments
This work was supported by PAPIIT--UNAM grant IN107706 to V.A. and CONACyT grant 42810 to
H.M.H.  J. Z. acknowledges support from DGEP-UNAM and CONACyT scholarships, and support by 
the CAS Research Fellowship for International Young Researchers. J. Z. is supported by 
the Joint Program in Astrophysical Cosmology of the Max Planck Institute for Astrophysics 
and the Shanghai Astronomical Observatory. We are grateful to S. Courteau for comments
to an early version of the manuscript and to the anonymous referee for his/her critizism
and suggestions.  


\vspace{0.5cm}
\centerline{APPENDIX A: ERROR BUDGET}
\vspace{0.3cm}

In the following we describe how we estimate the uncertainties for the
quantities we used to construct the scaling relations presented along the paper. We assume
them to be Gaussian distributed, so our estimates refer to the standard deviation.

{\it $B$ and $K$ magnitudes:} we neglect the contributions to the errors due to uncertainties 
on the redshift, the extinction in our Galaxy, the disk thickness, and the $k$ correction. 
Hence, the square of the error in the absolute magnitudes is estimated as the quadratic sum 
of the measurement error, $\epsilon_m$ (given in the original papers used to compile the 
sample), and the one due to the correction for internal extinction, $\epsilon_{A_i}$:
\begin{equation}
\epsilon_{M_{abs}}^2 = \epsilon_m^2 + \epsilon_{A_i}^2,
\label{mag}
\end{equation}
To estimate $\epsilon_{A_i}$ we take into account only errors related to the inclination of the 
galaxies, $\epsilon_{b/a}$, and neglect errors related to the empirical parameter $\gamma$ 
(recall 
that $A_i=\gamma(\W)log(b/a)$, Tully et al. 1998). Giovanelli et al. (1997) suggest to add
also the error $\epsilon_{\gamma}=0.15\gamma$, but this term is in most of the cases 
sub-dominant compared to the error in the inclination (see, for instance, fig. 2 of 
their paper); 
only for high inclination and high values of $\gamma$ both terms may become comparable. 
Thus, we have
\begin{equation}
\epsilon_{A_i}= (\frac{0.434}{b/a}\gamma)\epsilon_{b/a},
\end{equation}
where we assume $\epsilon_{b/a}= 0.09-0.12(1-b/a)+0.037(1-b/a)^2$, following the estimation 
for the median error of $\epsilon_{b/a}$ given by Giovanelli et al. (1997). Finally, the error 
in log$L$ is given by $\epsilon_{logL}=\epsilon_{M_{abs}}/2.5.$, where $\epsilon_{M_{abs}}$
was given in equation (\ref{mag}).

{\it Stellar mass:} The error in log\ms~ is estimated as the quadratic sum of the logarithmic 
errors in $L_K$ and in the mass--to--luminosity ratio $\Upsilon_{K}$:
\begin{equation}
\epsilon_{log\ms}^2 = \epsilon_{logL_K}^2 + \epsilon_{log\Upsilon_{K}}^2,
\end{equation}
where an uncertainty of 25\% in the value $\Upsilon_{K}$ was considered (Bell et al. 2003a).

{\it Baryonic mass:} The error is calculated as the quadratic sum of the errors in \ms~ and 
\mg, $\epsilon_{\md}^2 =  \epsilon_{\ms}^2 + \epsilon_{\mg}^2$;
the corresponding logarithmic error in \md~ is then 
\begin{equation}
\epsilon_{log\md} = \frac{0.434}{\md}\epsilon_{\md}.
\end{equation} 
The error in \mg~ is calculated as:
\begin{equation}
\epsilon_{\mg} = \mg\frac{\epsilon_{S_\nu}}{S_\nu},
\end{equation} 
where the measurement error $\epsilon_{S_\nu}$ is taken as reported in the HyperLEDA information
system.
  
{\it Scale lengths:} In order to estimate the fitting error in the scale lengths of the galaxies 
used here, we have experimented with the fit of the observed SB profiles by using both a 
marking--the--disk method and the bulge--to--disk decomposition method. We have seen that 
typical uncertainties in the determination of $R$ by both methods are of the order of $5-10\%$ 
(see also MacArthur et al. 2003; Graham 2002). There is a very weak systematical trend for 
larger lengths determined by the latter method relative to those determined by the 
former one from late to early morphological types. The uncertainty in $R$ also depends on the 
SB profile type and other specific details. Finally  we decided to assign a logarithmic error 
of $\epsilon_{logR}=0.05$ ($\approx 11\%$) 
to the scale lengths in both the $B$ and $K$ bands, and  $\epsilon_{logR}=0.06$ to the
baryonic case, where the gas disk also contributes to the final surface density profile.  

{\it Rotational velocities:} The errors (standard deviations) in log\W, $\epsilon_{log\W}$, 
were taken from the sources, namely Verheijen (1997), Verheijen \& Sancisi (2001), and the 
HyperLeda information system.

\vspace{0.4cm}
\centerline{APPENDIX B: THE SCALING RELATIONS OF}

\centerline{CDM HALOS AND DISK GALAXY FORMATION}
\vspace{0.3cm}

Distinct CDM halos --those not contained within larger halos-- can be characterized by
the so--called virial mass and radius, \mv~ and \rv, and by a maximum circular velocity,
\vmh.  Halos which are contained inside larger ones (sub--halos) commonly have radii and
masses smaller than \rv~ and \mv, because their growth is truncated or even reversed due 
to tidal stripping; \vmh~ is less affected by these effects. It is 
well known that the masses and maximum circular velocities of CDM halos, and even 
of sub-halos, correlate tightly as $\vmh\propto \mh^{a}$ with $a\approx 0.32-0.30$, 
where \mh~ is the virial or truncated mass of the halo (e.g., Navarro et al. 1997; 
Avila-Reese et al. 1998, 1999, 2005; Bullock et al. 2001a).  
The correlation is basically an imprint of the 
linear power spectrum of fluctuations, whose variance at the galaxy scales decreases very 
slowly (logarithmically) with the fluctuation mass. This rough scale invariance 
naturally produces the $\vmh$-$\mh$ relation for the collapsed CDM halos.  The complex 
non--linear non--spherical hierarchical halo assembly process introduces only minor 
deviations upon this relation as well as a scatter.  
The cosmological TFR has an intrinsic scatter, \sigVMh, mainly due to the 
stochastic nature of the mass assembly histories, which also produces a scatter 
in the halo concentration parameter $c$ (Avila-Reese et al. 1998,1999; FA-R).

For distinct halos, by definition $\rv\propto \mv^{1/3}$; sub-halos tend to follow a similar 
relation between \rh~ and \mh, but with some scatter and with a smaller proportionality 
coefficient.  Consequently, \vmh~ and \rh~ are also correlated, as $\vmh\propto \rh^{\gamma}$
with $\gamma\approx 1-1.25$, although extra scatter is introduced in this relation
due to the variation in \vmh~ for a given mass, as in the case of the $\vmh$-$\mh$ relation.

Disk galaxies are assumed to form inside the growing CDM halos from the trapped baryons. 
Therefore, the scaling relations of the halos are expected to be imprinted
in the baryonic and stellar scaling relations of disk galaxies. We describe below 
the main physical ingredients of the FA-R models (see for details Avila-Reese et al. 1998; FA-R;
Avila-Reese \& Firmani 2000). An extended Press--Schechter approach is used to generate 
the MAHs of the halos from the primordial density fluctuation field, and a generalized 
secondary infall model is applied to calculate the time--by--time virialization of the 
accreting mass shells. The evolution and structure of the $\Lambda$CDM halos calculated in this 
way agree well with results from cosmological N--body simulations (Avila-Reese et al. 1999). 
The halo mass shells are assumed to have aligned rotation axis with specific 
angular momentum given by $j_{sh}(t_v)=dJ(t_v)/dM_v(t_v)$,  where
$J=\lambda GM_v^{5/2}/\left| E\right| ^{1/2}$, $J$, M$_v$ and $E$ are
the total angular momentum, mass, and energy of the halo at the shell
virialization time $t_v$. The halo spin parameter, $\lambda_h$, is assumed to
be constant in time. As a result of the assembling of these
mass shells, a present day halo ends with an angular momentum distribution
close to the (universal) distribution measured by Bullock et al. (2001b) 
in N--body simulations. 

A fraction \fb~ of the mass of each shell is assumed
to cool down and form a disk layer in a dynamical time. The radial mass 
distribution of the layer is calculated by equating its specific angular
momentum to that of its final circular orbit in centrifugal equilibrium
(detailed angular momentum conservation is assumed). The superposition
of these layers form the disk, which tends to be steeper in the center 
and flatter at the periphery than the exponential law. The disk surface density 
distribution is mainly determined by the halo angular momentum distribution. 
For example, for a given halo of radius \rh~ and mass \mh~, the characteristic 
size of the baryonic disk (and its typical surface density), described in a first 
approximation by the scale length, \hb, is determined mainly by the halo spin 
parameter, $\lambda_h$.  The distribution of $\lambda_h$
found in analytical and numerical studies is well described by a log--normal function, 
whose median and dispersion almost do not depend on the halo mass. While we 
assume here that the pre-- and post--disk formation spin parameters are 
equal, $\lambda_h=\lambda_d$,
in most of our discussions, the quantity $\lambda$ refers to the post--disk formation spin 
parameter and it could deviate from $\lambda_h$ if the baryon angular momentum is redistributed 
(in the halo or inside the disk) or if not all the halo gas falls into the disk.

The gravitational interaction of disk and inner halo (important for estimating \vm)
is calculated using an extended adiabatic invariance formalism, which differs 
from the usual one in that we take into account the ellipticity of the orbits, i.e., 
the circular orbit assumption is relaxed.
The disk SF at a given radius (azimuthal symmetry is assumed) is triggered by 
the Toomre gas gravitational instability criterion and self--regulated by a 
vertical disk balance between the energy input due to SNe and the turbulent 
energy dissipation in the ISM. This physical prescription 
naturally yields a Schmidt law with an index $n\lesssim 2$, varying slightly 
along the disk. The SF efficiency depends on 
the gas surface density determined mainly by $\lambda$, and on the gas accretion
rate determined by the cosmological MAH. Finally, we estimate the mass
of the (pseudo)bulge as the inner disk mass where the Toomre 
stellar parameter indicates disk instability.

\vspace{0.4cm}
\centerline{APPENDIX C: PREDICTIONS FOR THE}
\centerline{BARYONIC SCALING RELATIONS}
\vspace{0.2cm}

\centerline{\it C.1.  The \vm-\md~ relation}

In order to pass from the cosmological TFR to the baryonic
one, we should pass (i) from the halo \vmh~ to \vm, and (ii) from the halo mass
\mh~ to \md. The latter is given simply by \md=\fb \mh, where \fb~ is the (central) galaxy 
mass fraction. The distribution of \fb~ and its possible dependence on \mh~
and other halo properties is not well known. However, indirect and direct pieces
of evidence show that \fb~ should be much smaller than the universal baryon fraction, 
$\fb<< \Omega_{\rm b}/\Omega_{\rm dm}$ (e.g., Mo et al. 1998; FA-R; Smith et al. 2001; 
Bell et al. 2003b; Jim\'enez, Verde \& Oh 2003; Pizagno et al. 2005; Hoekstra et al. 2005; 
Mandelbaum et al. 2006; Blanton, Geha \& West 2007). 
Regarding item (i), the formation of the disk inside the dark halo and the gravitational 
drag produced on it, redistributes the inner mass profile. The \vm/\vmh~ ratio increases 
with the disk surface density and the disk baryon fraction \fb. For a given \mh, the 
disk has a larger surface density for a smaller value of $\lambda$. Therefore, the relation 
between \vmh~ and \vm~ depends on both $\lambda$ and 
\fb: \vm/\vmh= G($\lambda$,\fb). The function G($\lambda$,\fb) has been approximated in
Zavala (2003) from models of disk formation inside CDM halos based on Mo et al. (1998) and 
FA-R (note that FA-R used an adiabatic contraction formalism generalized for 
elliptical orbits; 
as a result, the contraction is weaker than in the simpler case of circular orbits; see also 
Gnedin et al. 2004). If $\lambda$ and \fb~ do not depend on mass, then the function G is
almost independent of mass, and then the baryonic TFR is expected to have a slope similar to the
cosmological TFR.  On the other hand, due to the gravitational disk--halo coupling, as \fb~ 
decreases, G also decreases in such a way that the objects in the baryonic TFR diagram shift
nearly parallel to the cosmological TFR. The function G becomes more sensitive to $\lambda$ as 
\fb~ becomes larger; for higher values of \fb, a large variation in G is expected. This has 
deep implications for the scatter in the baryonic TFR.

The intrinsic scatter in the baryonic TFR has at least three sources: 
(i) the original scatter from the halo cosmological TFR, (ii) the scatter in \vm~ 
due to the dispersion in $\lambda$, and (iii) the scatter in both \vm~ and
\md~ due to the average value and dispersion of \fb. As mentioned above,
the latter produces a shift of the models nearly along the same
relation, in such a way that its effect on the baryonic TFR scatter is 
expected to be small for reasonably average values of \fb~ ($\lesssim 0.08$). 
The major contribution to the scatter comes from 
the dispersion in the spin parameter $\lambda$. For a given \fb~ and 
\vmh, different values of $\lambda$ produce disks of different surface densities, and
therefore different values of \vm; such an effect is accounted for by the function G. For
a given distribution of $\lambda$, the scatter due to $\lambda$, 
$\sigma_\lambda$, increases as \fb~ increases. 
The scatter also increases slightly if we allow for a dispersion in \fb~ around the 
mentioned values (Gnedin et al. 2007). 
The scatter in the halo \vmh-\mh~ relation, $\sigVMh$, arises because halos of 
a given mass that were formed through a rapid mass assembling process are more 
concentrated and 
have a larger \vmh~ than those that assembled slowly (Avila-Reese et al. 1998,1999). 
The magnitude of this scatter slightly decreases with mass and has a value
of $\sigVMh\approx 0.035$ for a $\sim 10^{12}\msunh$ halo according to 
cosmological numerical simulations (Avila-Reese et al. 1999,2005).
According to the FA-R models, the average scatter in the baryonic TFR
is \sigintrV$\approx 0.053$ (\fb=0.05=const.).
This value is a lower limit under the assumptions made. It 
could be smaller if the distribution of $\lambda$ is narrower
than the one in the CDM halos. This could happen if, for example, disk galaxies 
form only in a subset of CDM halos, biased to have $\lambda$ distribution narrower than
the overall sample (e.g., D'Onghia \& Navarro 2007; Dutton et al. 2007; Gnedin et al. 2007), 
or if the baryon spin parameter becomes smaller due to angular momentum transport processes; 
then the galaxies formed from the low end of the halo $\lambda$ distribution would no longer 
be a disk type.

\vspace{0.3cm}
\centerline{\it C.2. The \hb-\md~ relation}

According to the models of disk galaxy formation discussed above, for a 
given halo radius \rh, the typical size of the disk, for instance its scale length 
\hb, is mainly proportional to $\lambda  g(c)\rh$, where $g(c)$ is a function that 
depends weakly on the NFW halo concentration parameter $c$ (Mo et al. 1998), given
on its own by the halo MAH.
Recall that detailed angular momentum conservation was assumed.
The weak anti--correlation of $c$ with \mh~ introduces a very weak correlation in
\hb~ with \md.  Therefore, if the model parameters $\lambda$ and \fb~ ($\equiv \md/\mh$) 
do not depend on the mass, then the slope of the \hb-\md~ correlation is expected to be only 
slightly steeper than the slope of the \rh-\mh~ correlation, which is $\sim 0.33$ 
(Appendix B). However, the correlation is predicted to be highly scattered, with a 
systematical shift in the normalization due to $\lambda$ (recall that \hb~ is directly 
proportional to $\lambda$ while \md, in principle, is independent of $\lambda$), and also 
due to \fb~ and $c$.  The first two parameters determine the surface density of the disk; 
therefore, in the \hb-\md~ diagram, the scatter should be strongly correlated 
with the surface density, e.g. \sigd. Note that a dependence of $\lambda$ or 
\fb~ on mass would imply a 
change in the slope of the \hb-\md~ correlation. For example, if $\lambda$ anti--correlates 
with mass, then this slope will be shallower than that of the initial \rh-\mh~ correlation, 
but if \fb~ anti--correlates with mass, then the slope will be steeper. 

\vspace{0.3cm}
\centerline{\it C.3. The \vm-\hb~ relation}

In the context of the simple models discussed in this 
appendix, we have $\vm\propto G(\lambda,\fb)\vmh$ and $\hb\propto \lambda g(c)\rh$. 
Thus, if for the \LCDM~ halos $\vmh\propto \rh$, then one obtains that 
$\vm \propto (G[\lambda,\fb]/\lambda g[c]) \hb$. 
As mentioned above, $g(c)$ correlates very weakly with mass, and hence with radius. 
Then, if $\lambda$ and \fb~ are independent of mass, the expected \vm-\hb~ correlation should 
be slightly shallower than the cosmological one. On the other hand, if \fb~ correlates
(anti--correlates) with mass, then the slope becomes steeper (shallower).
The scatter around the \vm-\hb~ correlation is expected to be the largest one of the 
three scaling relations. Galaxies are scattered in the \vm-\hb~ diagram by their associated 
values of $\lambda$, \fb~ and $c$. Differences in each of these values spread the galaxies 
in a divergent way in both axes. Therefore, the scatter of the \vm-\hb~ relation
is expected to correlate significantly with both \sigd~ and galaxy color.  
Recall that in the \hb-\md~ diagram, galaxies are scattered by $\lambda$ and $c$ only 
along the \hb~ axis. In the \vm-\md~ diagram the spread produced by these parameters 
is also only along one axis (\vm), but furthermore, the spread produced by \fb~ in both 
axes is in such a way that galaxies move nearly along the main relation; that is why 
the TFR is the least scattered among the three relations.

\end{document}